\documentclass[twocolumn]{aastex61}

\usepackage{graphicx, txfonts}
\usepackage{url}
\usepackage{multirow}
\usepackage{natbib}

\newcommand{\dhp}{(D/H)$_{\rm P}$}
\newcommand{\yp}{Y$_{\rm P}$}
\newcommand{\neff}{$N_{\rm eff}$}
\newcommand{\obhh}{$\Omega_{\rm B,0}\,h^{2}$}
\newcommand{\zabs}{$z_{\rm abs}=2.52564$}
\newcommand{\dpg}{$d(p,\gamma)^{3}{\rm He}$}

\newcommand{\CII}{C\,\textsc{ii}}

\newcommand{\DI}{\textrm{D}\,\textsc{i}}

\newcommand{\FeII}{Fe\,\textsc{ii}}

\newcommand{\HI}{\textrm{H}\,\textsc{i}}

\newcommand{\HII}{\textrm{H}\,\textsc{ii}}

\newcommand{\Lya}{Ly$\alpha$}
\newcommand{\Lyb}{Ly$\beta$}

\newcommand{\NHI}{$N(\textrm{H}\,\textsc{i})$}

\newcommand{\OI}{O\,\textsc{i}}

\newcommand{\SiII}{Si\,\textsc{ii}}

\def\rahr{^{\rm h}}
\def\ramin{^{\rm m}}
\def\rasec{\!\!^{\rm s}}
\def\decdeg{^{\circ}}
\def\decmin{'}
\def\decsec{\!\!''}

\received{\today}
\revised{\today}
\accepted{\today}
\submitjournal{ApJ}

\shorttitle{primordial deuterium at one percent}
\shortauthors{Cooke, Pettini, \&\ Steidel}

\begin{document}

\title{One percent determination of the primordial deuterium abundance\footnote{Based on observations collected at the W.M. Keck Observatory
which is operated as a scientific partnership among the California Institute of 
Technology, the University of California and the National Aeronautics and Space 
Administration. The Observatory was made possible by the generous financial
support of the W.M. Keck Foundation.}}

\correspondingauthor{Ryan J. Cooke}
\email{ryan.j.cooke@durham.ac.uk}

\author[0000-0001-7653-5827]{Ryan J. Cooke}
\altaffiliation{Royal Society University Research Fellow}
\affiliation{Centre for Extragalactic Astronomy, Department of Physics, Durham University, South Road, Durham DH1 3LE, UK}

\author{Max Pettini}
\affiliation{Institute of Astronomy, Madingley Road, Cambridge CB3 0HA, UK}
\affiliation{Kavli Institute for Cosmology, Madingley Road, Cambridge CB3 0HA, UK}

\author{Charles C. Steidel}
\affiliation{California Institute of Technology, MS 249-17, Pasadena, CA 91125, USA}



\begin{abstract}
We report a reanalysis of a near-pristine absorption system, located at a redshift \zabs\ toward the quasar
Q1243$+$307, based on the combination of archival and new data obtained with the HIRES echelle
spectrograph on the Keck telescope. This absorption system, which has an oxygen abundance
[O/H]=$-2.769\pm0.028$ ($\simeq1/600$ of the solar abundance), is among the lowest metallicity
systems currently known where a precise measurement of the deuterium abundance
is afforded. Our detailed analysis of this system concludes, on the basis of eight \DI\ absorption lines,
that the deuterium abundance of this gas cloud is $\log_{10}\,({\rm D/H}) = -4.622\pm0.015$, which is in
very good agreement with the results previously reported by Kirkman et al., but with an improvement
on the precision of this single measurement by a factor of $\sim3.5$. Combining this new estimate with
our previous sample of six high precision and homogeneously analyzed D/H measurements, we
deduce that the primordial deuterium abundance is $\log_{10}\,({\rm D/H})_{\rm P} = -4.5974\pm0.0052$
or, expressed as a linear quantity, $10^{5}({\rm D/H})_{\rm P} = 2.527\pm0.030$; this value corresponds
to a one percent determination of the primordial deuterium abundance. Combining our result with a big bang nucleosynthesis (BBN)
calculation that uses the latest nuclear physics input, we find that the baryon density derived from BBN
agrees to within $2\sigma$ of the latest results from the \textit{Planck} cosmic microwave background data.
\end{abstract}

\keywords{cosmology: observations -- cosmology: theory -- primordial nucleosynthesis -- quasars: absorption lines -- quasars: individual: Q1243$+$307}



\section{Introduction}
\label{sec:intro}

Modern cosmology is described by just six model parameters,
all of which are known to within a few percent. This model provides a reliable
description of the universe from seconds after the big bang until the present
epoch. However, we know that the `Standard Model' of cosmology and particle
physics is incomplete. For example, we have no definitive description of dark
matter and dark energy, nor do we fully understand the properties of neutrinos.
New physics beyond the Standard Model may be exposed by measuring
the cosmological model parameters at high precision, and there are
many teams that are searching for this new physics by studying the
cosmic microwave background (CMB), weak and strong lensing,
and by observing standard candles and rulers, to name a few examples.

In recent years, there have also been several efforts to measure the
chemical abundances of the elements that were made during the first
minutes after the big bang, a process that is commonly referred
to as `big bang nucleosynthesis' (BBN; for a general review of the subject,
see \citealt{Ste07,Cyb16,MatKusKaj17}). The abundances of the primordial elements
--- which include the isotopes of hydrogen, helium, and lithium --- are
sensitive to the physics of the early universe, and are therefore a
tool that allows us to test the Standard Model. Moreover, measuring
the abundances of these primordial elements currently provides our
earliest test of the Standard Model.

In order to reliably measure the \textit{primordial} element abundances,
we must first identify environments that are as close as possible to being pristine,
and therefore still retain a primordial composition of the light elements.
The best available measurements of the primordial element abundances
come from different environments; conventionally, the mass fraction of $^{4}$He (\yp) is derived
from the emission lines of nearby \HII\ regions in metal-poor star forming
galaxies \citep{IzoThuGus14,AveOliSki15},\footnote{\yp\ can also be measured
from the small-scale CMB temperature fluctuations \citep{Efs15}, albeit with
lower precision.} while the
primordial $^{7}$Li abundance is determined from the atmospheres of very
metal-poor stars \citep{Asp06,Aok09,Mel10,Sbo10,Spi15}. At present, there
are no reliable measurements of the primordial $^{3}$He abundance; however,
with future facilities this measurement may become possible
\citep[several different techniques are described by][]{BanRooBal02,McQSwi09,Coo15}.

\begin{table*}
\begin{center}
    \caption{\textsc{Journal of Keck HIRES observational data used in this analysis}}
    \hspace{-0.6cm}\begin{tabular}{@{}llccccc}
    \hline
   \multicolumn{1}{l}{Date}
& \multicolumn{1}{c}{Principal}
& \multicolumn{1}{c}{Program}
& \multicolumn{1}{c}{HIRES}
& \multicolumn{1}{c}{$v_{\rm FWHM}$}
& \multicolumn{1}{c}{Wavelength}
& \multicolumn{1}{c}{Exposure}\\
   \multicolumn{1}{l}{}
& \multicolumn{1}{c}{Investigator}
& \multicolumn{1}{c}{ID}
& \multicolumn{1}{c}{Decker}
& \multicolumn{1}{c}{(km~s$^{-1}$)}
& \multicolumn{1}{c}{Range (\AA)}
& \multicolumn{1}{c}{Time (s)}\\
  \hline
1999 April 17,18  &  Tytler  &  U32H  &  C5  &  $7.99\pm0.02$  & 3190--4665 & 23,400 \\
2000 March 13,14  &  Tytler  &  U02H  &  C5  &  $7.99\pm0.02$  & 3190--4665 & 32,400 \\
2006 June 2  &  Prochaska  &  U152Hb  &  C1  &  $6.28\pm0.02$  & 3225--6085 & 3,600 \\
2016 March 31  &  Cooke  &  N162Hb  &  C1  &  $6.28\pm0.02$  & 3225--6085 & 14,400 \\
  \hline
    \end{tabular}
    \label{tab:obssummary}
\end{center}
\end{table*}

The only other primordial element that is accessible with current facilities is
deuterium, which can be measured using gas clouds that are seen in absorption
against the light of an unrelated background light source (typically, a quasar)
\citep{Ada76}. Although this technique was proposed
more than four decades ago, the first measurements were only achieved some
20 years later; even now, only a handful detections of the neutral deuterium (\DI)
absorption lines have been made
\citep{BurTyt98a,BurTyt98b,PetBow01,OMe01,Kir03,Cri04,OMe06,Pet08,FumOmePro11,PetCoo12,Not12,Coo14,Rie15,Coo16,Bal16,Rie17,Zav17}.
However, as discussed recently by \citet{Coo14}, absorption line systems that
have \HI\ column densities near the threshold of a damped \Lya\ system
(DLA; \NHI~$\simeq10^{20.3}~{\rm cm}^{-2}$)\footnote{In this paper,
we use the term `DLA' to represent any absorption line system with
\NHI~$>10^{20.3}~{\rm cm}^{-2}$
and `sub-DLA' for systems with $10^{19.0}~<~$\NHI$/{\rm cm}^{-2}~<~10^{20.3}$.}
are the most suitable environments
to precisely measure the primordial deuterium abundance, \dhp\ (see also
\citealt{Rie17}). In this \HI\ column density regime, the \HI\ \Lya\ transition
exhibits Lorentzian damped wings that uniquely determine the total \HI\
column density, while up to $\sim10$ high order unsaturated \DI\ lines are
available to determine the total \DI\ column density. Even among DLAs, only
those that are kinematically quiescent are able to deliver a precise determination
of the primordial deuterium abundance, \dhp, since the \DI\ lines need to be
optically thin and unblended with nearby absorption lines.
Empirically, it has been noted by several authors that DLAs
with simple kinematics tend to be more common at the lowest
metallicity \citep{Led06,Mur07,Pro08,JorMurTho13,Nel13,CooPetJor15}.
Currently, there are just six systems which satisfy the above
conditions, all of which have been homogeneously analyzed, as reported in
previous papers of this series \citep{Coo14,Coo16}.

The primordial deuterium abundance, \dhp, is inferred under the assumption
that the ratio of deuterium to hydrogen atoms, ${\rm D/H}~\equiv~N$(\DI)/$N$(\HI).
There are several physical processes that potentially weaken
the validity of this assumption, including:
(1) The astration of deuterium as gas is cycled through generations of stars
\citep[][and the comprehensive list of references provided by \citealt{Cyb16}]{Dvo16,vdV17};
(2) the relative ionization of deuterium and hydrogen in neutral gas \citep{Sav02,CooPet16}; and
(3) the preferential depletion of deuterium onto dust grains \citep{Jur82,Dra04,Dra06}.
The first two physical processes are expected to alter the measured D/H ratio
by $\lesssim0.1$ per cent when the metallicity is $\lesssim 1/100$ solar and the
neutral hydrogen column density exceeds $10^{19}~{\rm cm}^{-2}$; this correction
is an order of magnitude below the
current measurement precision. The preferential depletion of deuterium onto
dust grains, however, has not been modeled in detail in metal-poor DLAs.

Several studies have reported on the depletion of deuterium in the local interstellar
medium (ISM) of the Milky Way \citep{Woo04,ProTriHow05,Lin06,EllProLop07,LalHebWal08,ProSteFie10}.
However, the ISM of the Milky Way is relatively dust-rich compared with the metal-poor
DLAs that are typically used to infer the primordial deuterium abundance. Observationally,
metal-poor DLAs are not expected to contain a significant amount of dust \citep{MurBer16};
even the most refractory elements in the lowest-metallicity DLAs are hardly incorporated
into dust grains \citep{Pet97,Vla04,Ake05}. However, \citet{Coo16} noted a subtle
(but statistically insignificant) decline of the deuterium abundance with increasing
metallicity, a trend that would be expected if deuterium were preferentially incorporated into
dust grains.

Herein, we report a seventh high precision measurement of the deuterium abundance
in one of the most pristine environments currently known, to assess whether or not
the deuterium abundance depends on metallicity. The paper is organized as
follows: In Section \ref{sec:obs}, we describe the observational procedure and the
details of the data reduction process. The analysis technique and the properties of
the absorption system are then described in Section \ref{sec:analysis}.
In Section~\ref{sec:sample}, we report our new D/H abundance
measurement of this system, and investigate the properties of our full sample.
In Section~\ref{sec:cosmology}, we deduce the primordial deuterium abundance, based
on seven D/H values, and provide new measurements of the cosmological baryon density,
and effective number of neutrino species. Our conclusions are summarized in
Section~\ref{sec:conc}. All reported uncertainties represent $68\%$  confidence
intervals, unless otherwise stated.

\section{Observations and Data Reduction}
\label{sec:obs}

\subsection{Observational Data}
\label{sec:obsdata}

This paper presents an estimate of the primordial D/H abundance using new, high quality
data of a previously known sub-DLA at an absorber redshift
$z_{\rm abs}\simeq2.5257$ toward the quasar Q1243$+$307 ($z_{\rm em}\simeq2.558$,
Right Ascension~$=12\rahr46\ramin10.\rasec9$,
Declination~$=+30\decdeg31\decmin31.\decsec2$; J2000). A measure of the deuterium
abundance of this system was first reported by \citet{Kir03}, using data taken with the
High Resolution Echelle Spectrometer (HIRES; \citealt{Vog94}) on the Keck I telescope
during the years $1999-2000$ (program IDs: U32H, U02H). These data consist of a total exposure time of 55,800\,s,
divided into seven exposures, acquired with the previous generation HIRES detector;
this detector had relatively low UV quantum efficiency and significantly higher read noise
at the bluest wavelengths where the redshifted
\DI\ absorption lines are observed. For these reasons,\footnote{This system was not analyzed in our
previous work \citep{Coo14}, since the data were not publicly available at the time.} we
have re-observed Q1243$+$307 using the modern HIRES detector,
which is considerably more sensitive at blue wavelengths.

Our observations (program ID: N162Hb) consisted of $3\times3600$~s and $1\times3000$~s exposures,
and were carried out on 2016 March 30, in excellent seeing conditions
($0.6''$ full width at half maximum; FWHM), well matched to the chosen slit size (C1 decker, $0.861''\times7.0''$).
This decker provides a nominal spectral resolution of $R\simeq48,000$ ($v_{\rm FWHM}\simeq6.25$~km~s$^{-1}$)
for a uniformly illuminated slit. Using an exposure of a thorium--argon (ThAr) lamp, we directly measured the
instrument FWHM to be $v_{\rm FWHM}=6.28\pm0.02$~km~s$^{-1}$ based on 2192 emission lines;
throughout our analysis, we adopt this FWHM value.\footnote{Ideally, the instrument FWHM should be
determined using narrow telluric absorption lines, since the quasar was not uniformly illuminating the slit
during the observations. Unfortunately, there are no telluric absorption bands covered by our spectrum,
and we have therefore adopted the FWHM value of a uniformly illuminated slit. We note that this
assumption should not affect our determination of \DI/\HI, because the equivalent width of
an absorption line is invariant under convolution with the instrumental FWHM. As discussed in
Section~\ref{sec:analysis}, the equivalent widths of the weak \DI\ absorption lines and the damped
profile of the strong \Lya\ absorption line uniquely determine the \DI\ and \HI\ column densities, respectively.}
All science and calibration frames were
binned $2\times2$ during read-out. The final combined signal-to-noise ratio (S/N) per 2.5~km~s$^{-1}$
pixel of our data near the observed wavelength $\lambda_{\rm obs} = 3215$\,\AA\ (i.e. the Lyman limit of the sub-DLA)
is S/N~$\simeq~25$. The S/N ratio is much higher at longer wavelengths, and reaches a maximum value of
S/N~$\simeq~80$ per 2.5~km~s$^{-1}$ pixel near the sub-DLA's \Lya\ absorption line.

Finally, a single exposure of length 3600\,s was acquired with Keck+HIRES on
2006, June 2 (program ID: U152Hb, \citealt{Leh14,OMe15}), using a nearly identical setup as our own
observations (hereafter referred to as the KODIAQ data). We have retrieved all of the aforementioned
HIRES data of Q1243$+$307 from the public Keck Observatory
data archive.\footnote{Available from: \url{https://koa.ipac.caltech.edu/cgi-bin/KOA/nph-KOAlogin}}
A summary of the data that are used in our analysis is provided in Table~\ref{tab:obssummary}.

\subsection{Data Reduction Methods}

The modern HIRES data (program IDs: U152Hb, N162Hb) were reduced using the
HIRES Redux package \citep{BerBurPro15}.\footnote{HIRES Redux is available from: \url{http://www.ucolick.org/~xavier/HIRedux/}} We adopted the standard processing steps, including a bias level
subtraction, correcting for the pixel-to-pixel variations and dividing by the blaze
function of each echelle order. The echelle orders were traced using an exposure of a
quartz lamp taken through the C1 decker (i.e. the same slit that was used to acquire the
science frames). We employed an optimal sky subtraction and object extraction
technique \citep{Kel03}, and each pixel was assigned a wavelength using an
exposure of a ThAr lamp that bracketed each science exposure.

At the time of our analysis, the HIRES Redux package was not able to reduce data acquired with
the old HIRES detector (see, however, \citealt{OMe17}). We therefore reduced the \citet{Kir03} data (program IDs: U32H, U02H) using version 5.2.4 of the \textsc{makee} data
reduction pipeline,\footnote{\textsc{makee} is available from:
\url{http://www.astro.caltech.edu/~tb/makee/}} adopting a similar approach as that described by \citet{Suz03}.
To summarize, we performed a bias subtraction, a flatfield and blaze correction,
and traced the orders using an exposure of a quartz lamp taken through a pinhole decker.
The data were wavelength calibrated using a ThAr lamp; we measured the widths of
1040 ThAr lines to be $v_{\rm FWHM}=7.99\pm0.02$~km~s$^{-1}$, which is in good
agreement with the nominal value of the instrument resolution ($v_{\rm FWHM}=8.0$~km~s$^{-1}$).
For the analysis of the \citet{Kir03} dataset described below, we adopt our measured
value of the FWHM.

\begin{figure*}
  \centering
 {\includegraphics[angle=0,width=155mm]{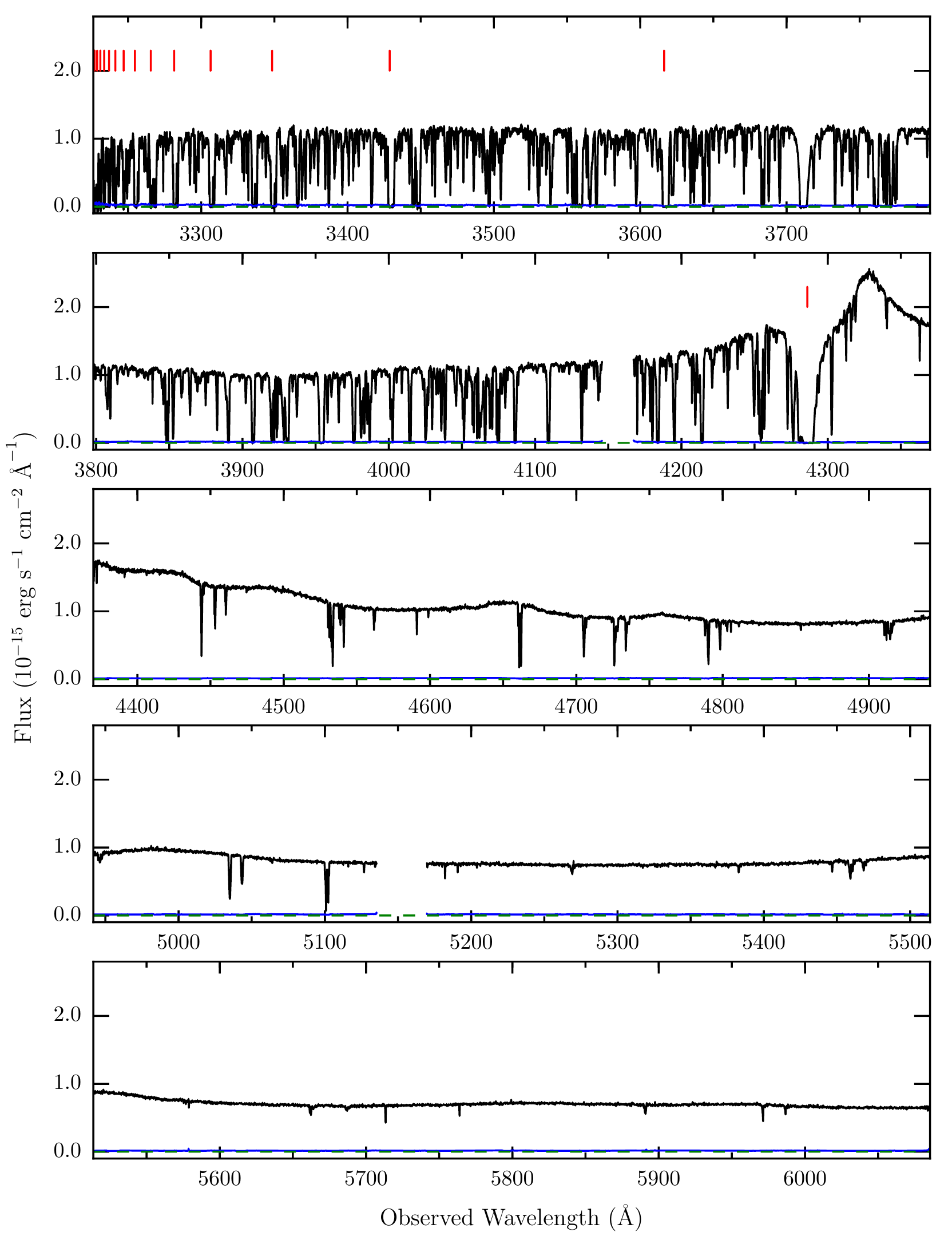}}\\
  \caption{
The final combined and flux-calibrated spectrum of Q1243$+$307 is shown (black histogram)
with the corresponding error spectrum (blue histogram) and zero level (green dashed line).
The red tick marks above the spectrum indicate the locations of the Lyman series absorption
lines of the sub-DLA at redshift \zabs. Note the exquisite signal-to-noise ratio of the combined 
spectrum, which varies from S/N~$\simeq~80$ near the \Lya\ absorption line of the sub-DLA ($\sim4300$\,\AA)
to  S/N~$\simeq~25$ at the Lyman limit of the sub-DLA, near 3215\,\AA\ in the observed frame.
}
  \label{fig:spectrum}
\end{figure*}

All reduced data were corrected to the heliocentric frame of reference, and were
converted to a vacuum wavelength scale. Using \textsc{UVES\_popler},\footnote{\textsc{uves\_popler}
is maintained by Michael T. Murphy, and is available from GitHub, via the following
link: \url{https://github.com/MTMurphy77/UVES\_popler}} we combined the exposures of each given
dataset to produce three separate spectra of Q1243$+$307; one spectrum of the
\citet{Kir03} data, one of the KODIAQ data, and the combined spectrum of our new data.
As described in Section~\ref{sec:analysis}, all three of these spectra are kept separate
from one another, but are analyzed simultaneously.
For illustration purposes, in Figure~\ref{fig:spectrum} we show the complete combined
spectrum of our new data (i.e. only the data acquired in 2016), flux calibrated
with reference to the \citet{Kir03} data.

\section{Analysis Method}
\label{sec:analysis}

\begin{figure*}
  \centering
 {\includegraphics[angle=0,width=140mm]{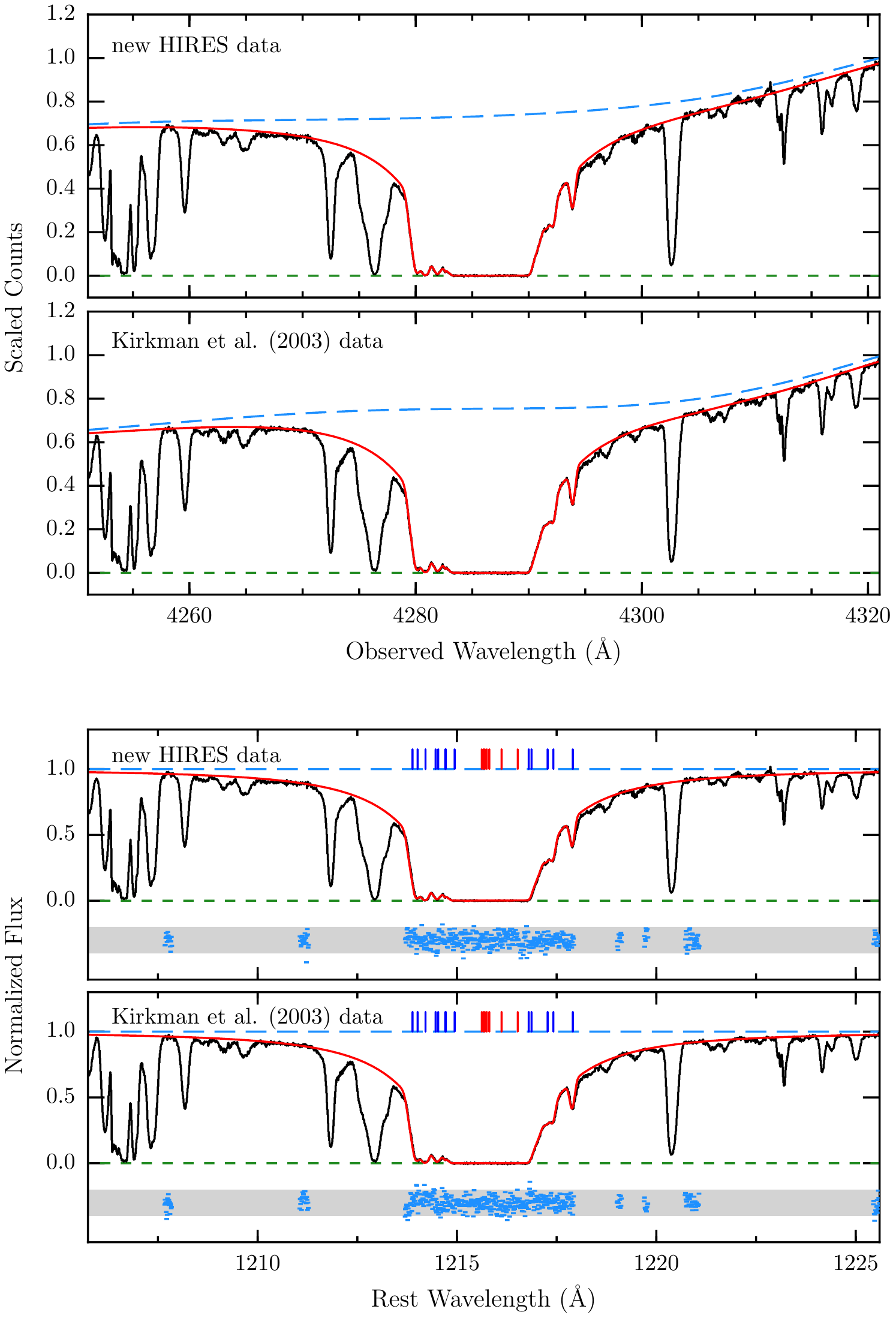}}\\
  \caption{
The \Lya\ profile of the absorption system at \zabs\ towards the quasar Q1243$+$307
is shown (black histogram) overlaid with the best-fitting model profile (red line), continuum (long dashed blue line),
and zero-level (short dashed green line). The top panels show the raw, extracted counts scaled to the maximum
value of the best-fitting continuum model. The bottom panels show the continuum normalized flux spectrum.
The label provided in the top left corner of every panel indicates the source of the data. The blue points
show the normalized fit residuals, (data--model)/error, of all pixels used in the analysis, and the gray
band represents a confidence interval of $\pm2\sigma$. The signal-to-noise
ratio is comparable between the two datasets at this wavelength range, but it is markedly different near
the high order Lyman series lines (see Figures~\ref{fig:lyseriesa} and \ref{fig:lyseriesb}).
The red tick marks above the spectra in the bottom panels show the absorption components
associated with the main gas cloud (Components 2, 3, 4, 5, 6, 8, and 10 in Table~\ref{tab:compstruct}),
while the blue tick marks indicate the fitted blends. Note that some blends are also detected in Ly$\beta$--Ly$\epsilon$.
}
  \label{fig:lya}
\end{figure*}

We now summarize the main aspects of our analysis method, which is
identical to that described in our previous work \citep{Coo14,Coo16}.
We use the Absorption LIne Software (\textsc{alis}), which employs a
$\chi$-squared minimization procedure to minimize the residuals between
the data and a user-specified model, weighted by the inverse variance of the
data.\footnote{\textsc{alis} is available for download from GitHub: \url{https://github.com/rcooke-ast/ALIS}}

A key aspect of our analysis is that we simultaneously fit the emission spectrum of the quasar and the absorption
due to the intervening absorption line system. This approach ensures that the final error on D/H
includes the uncertainty associated with the quasar continuum placement.
We include all available information of the
absorption system in our analysis, including the \HI\ and \DI\ Lyman series
absorption lines and the unblended metal absorption lines. The continuum
near each absorption line is fit during the minimization process
assuming that it is described by a low order Legendre polynomial.
Typically, the degree of the Legendre polynomial is $\lesssim4$, except near
the \Lya\ absorption line, where a Legendre polynomial of degree 8
is used. We also include a global model parameter that defines the zero-level
of each dataset, to account for small residuals in the background subtraction
and/or partial covering of the background quasar by the foreground gas cloud.
All three datasets are analyzed at the same time to obtain a global best-fit model;
we simultaneously fit the same absorption model to all three datasets,
while allowing the model of the quasar continuum around every absorption
line in each dataset to be different.\footnote{There are two reasons why the
emission profile near each absorption line may be different for the three
datasets. First, the three datasets that are analyzed in this paper were taken at different
epochs; the quasar continuum and emission lines may vary over the 16-year period
spanned by the observations.
Second, these datasets were acquired with different instrument configurations;
the relative differences in the spectrograph efficiency as a function of wavelength
can change the `apparent' level of the quasar continuum.} We also
include two fitting parameters to determine the global relative velocity shift
between the three datasets, to account for instrumental artifacts in the
wavelength calibration \citep[e.g.][]{WhiMur15}.

\begin{figure*}
  \centering
 {\includegraphics[angle=0,width=160mm]{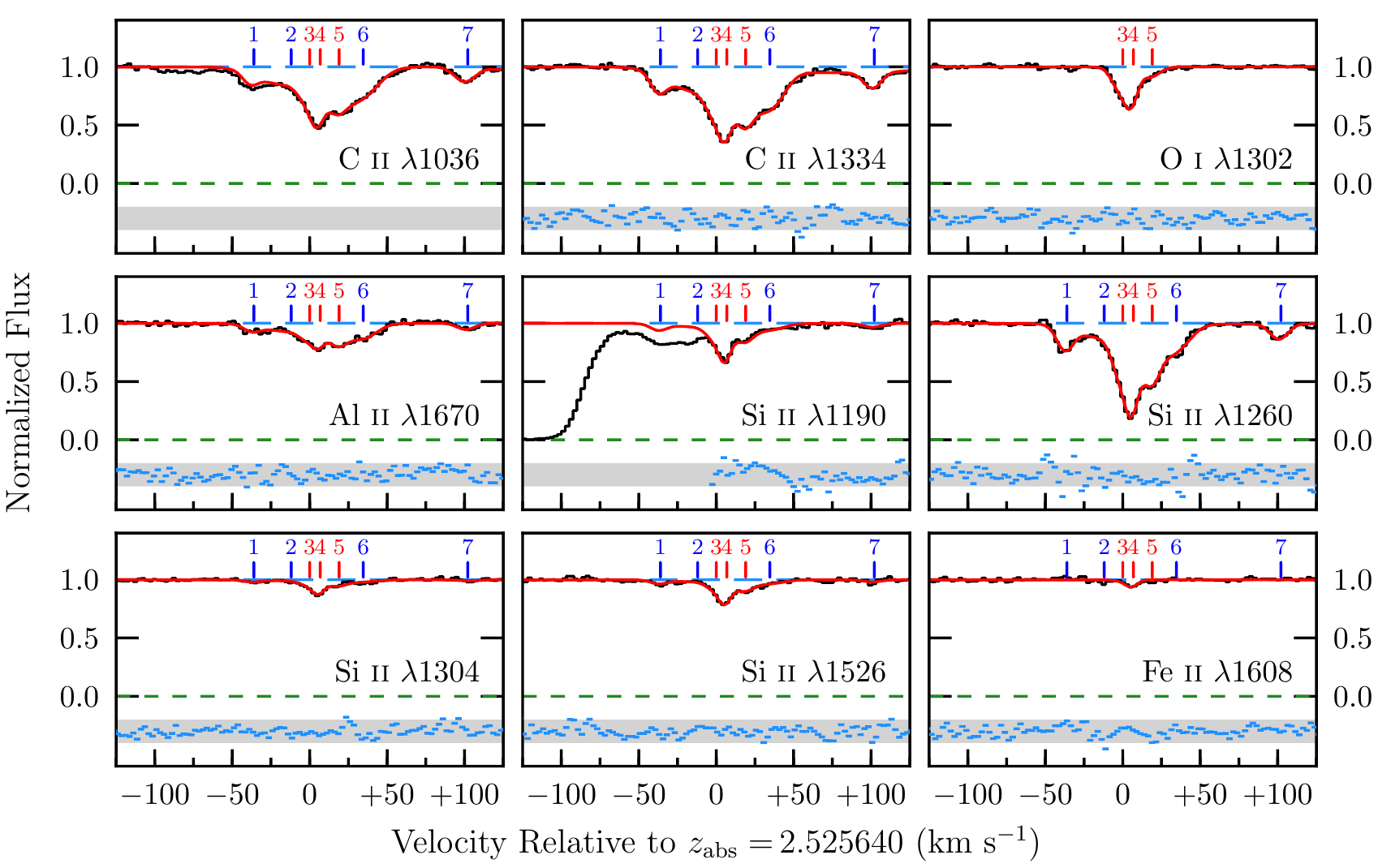}}\\
  \caption{The metal absorption lines that are used in our analysis are shown as a black histogram,
  overlaid with the best-fitting model (red line). The data and model are normalized to the best-fitting
  continuum model (long dashed blue line) and corrected for the fitted zero-level (short dashed green
  line). The red tick marks above each spectrum indicate the location of the absorption components
  seen in neutral gas (Components 3, 4, 5; see Table~\ref{tab:compstruct}), while the blue tick marks indicate the absorption components
  that are only seen in ionized gas (remaining components; see Table~\ref{tab:compstruct}). The number above each tick mark indicates the Component Number, which is listed in the first column of Table~\ref{tab:compstruct}.
  The blue points below each spectrum are the normalized fit residuals, (data--model)/error, of all pixels used in the analysis, and the gray band represents a confidence interval of $\pm2\sigma$. 
  The different profiles exhibited by the neutral (\OI) and single ionized species (all remaining ions shown) are likely to be the
  result of ionized gas.}
  \label{fig:metals}
\end{figure*}

\begin{deluxetable*}{@{}crrrcccccccc}
\tablecaption{Best-fitting model parameters of the absorption system at \zabs\ towards the QSO Q1243$+$307$^{\rm d}$\label{tab:compstruct}}
\tablewidth{700pt}
\tabletypesize{\scriptsize}
\tablehead{
\colhead{Component} &
\colhead{$z_{\rm abs}$}  &
\colhead{$\Delta v$}  &
\colhead{$b_{\rm turb}$}  &
\colhead{$\log_{10} N$\/(H\,{\sc i})} &
\colhead{$\log_{10} {\rm (D\,\textsc{i}/H\,\textsc{i})}$} &
\colhead{$\log_{10} N$\/(C\,{\sc ii})} &
\colhead{$\log_{10} N$\/(O\,{\sc i})} &
\colhead{$\log_{10} N$\/(Al\,{\sc ii})} &
\colhead{$\log_{10} N$\/(Si\,{\sc ii})} &
\colhead{$\log_{10} N$\/(S\,{\sc ii})} &
\colhead{$\log_{10} N$\/(Fe\,{\sc ii})} \\
\colhead{Number}  &  \colhead{}  &
\colhead{(km~s$^{-1}$)} &
\colhead{(km~s$^{-1}$)} &
\colhead{(cm$^{-2}$)} &
\colhead{} &
\colhead{(cm$^{-2}$)} &
\colhead{(cm$^{-2}$)} &
\colhead{(cm$^{-2}$)} &
\colhead{(cm$^{-2}$)} &
\colhead{(cm$^{-2}$)} &
\colhead{(cm$^{-2}$)} 
} 
\startdata
1   & $2.525216$                &   $-36.1$     & $5.9$             & \nodata$^{\rm b}$  & $-4.622^{\rm a}$         &  $12.70$                  &  \nodata$^{\rm b}$   & $10.96$                  &  $11.87$                &  \nodata$^{\rm b}$ &  \nodata$^{\rm b}$   \\  
     &  $\pm 0.000002$        &   $\pm0.9$      &$\pm 0.4$       &                               & $\pm 0.015$                &  $\pm  0.04$            &                                 & $\pm 0.09$             & $\pm 0.02$            &                               &                                  \\
2   & $2.52550$                  &   $-11.9$      & $23.0$           & $17.23$                 & $-4.622^{\rm a}$         &  $13.32$                  &  \nodata$^{\rm b}$   &  $11.60$                 &  $12.14$                 &  \nodata$^{\rm b}$ &  \nodata$^{\rm b}$   \\  
     &  $\pm 0.00003$          &   $\pm2.7$      &$\pm 0.8$       & $\pm 0.08$            & $\pm 0.015$                &  $\pm  0.04$            &                                 &  $\pm 0.05$            &  $\pm 0.04$           &                               &                                  \\
3   & $2.52564$                  &   $0.0$      & $5.8$             & $19.58$                 & $-4.622^{\rm a}$         &  $13.01$                  &  $13.27$                 &  $11.22$                 &  $12.23$                 &  \nodata$^{\rm b}$ &  $11.73$                  \\  
     &  $\pm 0.00001$          &   $\ldots$      &$\pm 0.9$       & $\pm 0.07$            & $\pm 0.015$                &  $\pm  0.11$            &   $\pm 0.08$           &   $\pm 0.12$           &  $\pm 0.10$            &                               &   $\pm 0.39$            \\
4   & $2.525720$               &    $+6.8$      & $3.4$             & $19.08$                 & $-4.622^{\rm a}$         &  $13.21$                  &   $13.31$                 &  $11.28$                 &  $12.55$                &  $12.59$                 &  $12.39$                  \\  
     &  $\pm 0.000002$       &   $\pm0.9$       &$\pm 0.3$       & $\pm 0.23$            & $\pm 0.015$                &  $\pm  0.06$            &  $\pm 0.08$             &  $\pm 0.09$            &  $\pm 0.05$          &  $\pm 0.12$            &   $\pm 0.07$             \\
5   & $2.525864$              &     $+19.0$      & $7.3$             & $18.68$                 & $-4.622^{\rm a}$         &  $13.28$                  &  $12.91$                  &  $11.53$                  &  $12.41$                &  \nodata$^{\rm b}$ &  \nodata$^{\rm b}$   \\  
     &  $\pm 0.000003$      &    $\pm0.9$       &$\pm 0.4$       & $\pm 0.17$            & $\pm 0.015$                &  $\pm  0.04$            &  $\pm 0.03$             &  $\pm 0.05$            &    $\pm 0.03$          &                               &                                  \\
6   & $2.526047$              &   $+34.6$        & $12.3$           & $16.86$                 & $-4.622^{\rm a}$         &  $13.34$                  &  \nodata$^{\rm b}$   &  $11.64$                  &  $12.19$                &  \nodata$^{\rm b}$ &  \nodata$^{\rm b}$   \\  
     &  $\pm 0.000007$       &  $\pm1.0$        &$\pm 0.6$       & $\pm 0.04$            & $\pm 0.015$                &  $\pm  0.03$            &                                 &   $\pm 0.04$           &   $\pm 0.03$           &                               &                                  \\
7   & $2.526841$              &    $+102.1$       & $6.8$             & \nodata$^{\rm b}$  & $-4.622^{\rm a}$         &  $12.69$                   &  \nodata$^{\rm b}$   &  $11.01$                 &  $11.71$                &  \nodata$^{\rm b}$ &  \nodata$^{\rm b}$   \\  
     &  $\pm 0.000003$      &  $\pm0.9$         &$\pm 0.4$       &                               & $\pm 0.015$                &  $\pm  0.04$             &                                 &  $\pm 0.05$            &   $\pm 0.02$          &                               &                                  \\
8   & $2.526943$              &   $+110.8$        & $25.9$           & $16.242$               & $-4.622^{\rm a}$         &  \nodata$^{\rm b}$   &  \nodata$^{\rm b}$   &  \nodata$^{\rm b}$  &  \nodata$^{\rm b}$ &  \nodata$^{\rm b}$ &  \nodata$^{\rm b}$   \\  
     &  $\pm 0.000003$      &   $\pm0.9$        &$\pm 0.2$       & $\pm 0.004$          & $\pm 0.015$                &                                 &                                 &                                &                                &                               &                                  \\
9 & $2.52797$               &     $+198.1$       & $8.3$             & \nodata$^{\rm b}$  & $-4.622^{\rm a}$         &  $11.57$                   &  \nodata$^{\rm b}$   &  $10.78$                  &  $10.82$               &  \nodata$^{\rm b}$ &  \nodata$^{\rm b}$   \\  
     &  $\pm 0.00002$       &    $\pm1.9$        &$\pm 2.6$       &                               & $\pm 0.015$                &  $\pm  0.36$             &                                 &   $\pm 0.13$            &  $\pm 0.14$          &                               &                                  \\
10 & $2.528115$             &     $+210.5$        & $27.6$           & $16.365$               & $-4.622^{\rm a}$         &  \nodata$^{\rm b}$    &  \nodata$^{\rm b}$   &  \nodata$^{\rm b}$ &  \nodata$^{\rm b}$ &  \nodata$^{\rm b}$ &  \nodata$^{\rm b}$   \\  
     &  $\pm 0.000002$     &   $\pm0.9$         &$\pm 0.1$       & $\pm 0.004$          & $\pm 0.015$                &                                  &                                 &                               &                                &                               &                                  \\
\hline
Total$^{\rm c}$ & & &  & $19.761$               & $-4.622^{\rm a}$         &  $13.663$                   &  $13.681$                  &  $11.850$                  &  $12.900$                 &  $12.59$                       &  $12.507$                         \\
                        &  & &  & $\pm 0.026$          & $\pm 0.015$                &  $\pm  0.024$            &  $\pm  0.011$              & $\pm  0.033$            &$\pm  0.011$             &$\pm  0.12$            & $\pm0.085$                                 \\
\enddata
\tablenotetext{\rm a}{Forced to be the same for all components.}
\tablenotetext{\rm b}{Absorption is undetected for this ion in this component.}
\tablenotetext{\rm c}{The total column densities quoted are those in Components 3, 4, and 5, which together account for $99.5\%$ of the total column density of neutral gas in this system. We note that the individual
\HI\ component column densities are strongly degenerate with each
other due to the multi-component structure of the sub-DLA. Although
the total \HI\ column density is largely unaffected, the reported uncertainty
of $N$\,(H\,{\sc i}) reported here is likely overestimated. A more accurate estimate
of the uncertainty on the total \HI\ column density would be afforded by fitting
directly to the total \HI\ column density during the chi-squared
minimization, but this feature is not yet implemented in \textsc{alis}
when the \DI/\HI\ ratio is forced to be the same in every component.}
\tablenotetext{\rm d}{The resulting $\chi$-squared/dof of this model is $12115/13697\simeq0.885$ (see footnote~\ref{foot:chisq}).}

\tablecomments{By comparing the relative widths of the metal and \DI\ absorption lines,
we determine the kinetic temperature of Component 3 and Component 4 to be
$T_{\rm kin}=8820\pm820~{\rm K}$ and $T_{\rm kin}=4100\pm2300~{\rm K}$, respectively.
For Components 5 and 6, we allowed the total Doppler parameters of the \HI\ Lyman series
absorption lines to vary independently of the metal absorption lines, which is equivalent to
adding a thermal contribution to the line profiles. The resulting total Doppler parameters of
the \HI\ lines of Components 5 and 6 are
$b_{\rm tot}=14.6\pm0.6~{\rm km~s}^{-1}$ and $b_{\rm tot}=24.0\pm0.7~{\rm km~s}^{-1}$.
The remaining absorption components are not sensitive to the relative contributions of
turbulent and thermal broadening.}
\end{deluxetable*}

As discussed in the Introduction, accurate estimates of the primordial deuterium
abundance are afforded by systems where the wings of the \Lya\ absorption line
are damped by the Lorentzian term of the Voigt profile. In this regime, the damped
wings uniquely determine the \HI\ column density. For this reason, the data are
most sensitive to \NHI\ when the optical depth of the absorption profile is
$\tau\gtrsim0.7$ (i.e. where the residual intensity is $\lesssim50$~per~cent of the
quasar continuum); for the sub-DLA towards Q1243$+$307, this corresponds
to all pixels within $\pm470~{\rm km~s^{-1}}$ of the \HI\ \Lya\ line relative to the
redshift defined by the narrow metal absorption lines. In our analysis,
we opted to include all pixels that are
within a velocity interval of $-470\le v/{\rm km~s^{-1}}\le+580$ (see also, Section~\ref{sec:senslya}).
Any blends that are identified
within this velocity interval are modeled with a Voigt profile; outside this
velocity interval we only include pixels in the $\chi$-squared minimization that
are deemed by visual inspection to be free of unrelated absorption. We present the
data and best-fitting model profile of the \Lya\ absorption feature in
Figure~\ref{fig:lya}. The best-fitting model profile has an \HI\ column
density of $\log_{10} N$\,(H\,{\sc i})/cm$^{-2}=19.761\pm0.026$, which is
in good agreement with the corresponding estimate reported by
\citet{Kir03}, $19.73\pm0.04$.

In Figure~\ref{fig:metals}, we present the metal absorption lines in the \zabs\ sub-DLA
that were used in our analysis. We only show the new Keck HIRES data in this figure,
but we note that our analysis includes all of the data that were described in
Section~\ref{sec:obsdata}.
It is immediately obvious that the component structure of the metal absorption lines
is complex, with the neutral species (e.g., \OI) exhibiting a different
structure to the singly ionized species (e.g., \CII). This difference is probably
due to the presence of ionized gas in the absorption system (see also, Section~\ref{sec:sensmetals}).
Table~\ref{tab:compstruct} lists full details of our absorption profile
model which we now briefly summarize.

We assume that the gas in each absorption component is distributed according
to a Maxwell-Boltzmann distribution, such that every component is represented by
a Voigt profile characterized by a column density, a total Doppler parameter, and a
redshift. We model the total Doppler parameter with a contribution from turbulent
and thermal broadening:
\begin{equation}
b_{\rm total}^{2} = b_{\rm turb}^{2} + b_{\rm therm}^{2} \equiv b_{\rm turb}^{2} + 2\,k_{\rm B}\,T_{\rm gas}/m_{\rm ion}
\end{equation}
where $T_{\rm gas}$ is the gas temperature, $m_{\rm ion}$ is the mass of the
ion responsible for the absorption line, and $k_{\rm B}$ is the Boltzmann constant.
As noted previously by \citet{Coo14}, at the high quality of the data typically
acquired for D/H measurements, this model is too simplistic; in reality, there is a
distribution of turbulence and gas temperature along the line-of-sight. To overcome this
simplicity, we assume that all gas constituents share the same redshift and turbulent
Doppler parameter, while the \DI\ and \HI\ thermal broadening components are fit separately. This
prescription offers enough flexibility so that the total Doppler parameter of the \HI, \DI,
and metal absorption lines can be determined almost independently. However,
we emphasize that
the weak \DI\ absorption lines and the strong \HI\ damped \Lya\ line do not depend on our
choice to model the absorption lines as a Voigt profile. The \DI\ column density
depends only the equivalent widths of several weak absorption lines, while the Lorentzian
damped \HI\ \Lya\ line uniquely determines the \HI\ column density. The resulting
\DI/\HI\ ratio is therefore unaffected by this assumption.

We find that the neutral \OI\ absorption can be accurately described by three
model components (denoted Component 3, 4, and 5 in Table~\ref{tab:compstruct}),
which are indicated by the red tick marks above the spectrum in Figure~\ref{fig:metals}.
An additional five components are required to accurately represent the
component structure of the ionized gas (Component 1, 2, 6, 7, and 9), shown as blue tick marks
above the spectra in Figure~\ref{fig:metals}.\footnote{In order to emphasize
the structure of the absorption profile around the neutral absorption components,
we have not shown the absorption of Component 9
(located at a velocity of $\Delta v=+198.1~{\rm km~s}^{-1}$
relative to Component 3) in Figure~\ref{fig:metals}, since it is very weak.}
Finally, there are an additional two components
that are seen in the \HI\ gas that are not seen in the low ion metal absorption
lines (Components 8 and 10); these are also the two weakest \HI\ components that are detected.
We also point out that three of the absorption components (Components 1, 7, and 9) are
detected in the metal absorption lines and do not have a discernible \HI\ column
density; it is difficult to resolve the \HI\ absorption in these components due to
their proximity to the stronger \HI\ absorption exhibited by Components 3, 4, and 5.

Due to the presence of ionized gas, we only provide an estimate of the oxygen
abundance of this absorption system; $N$(\OI)/$N$(\HI) is considered a reliable
measure of the [O/H] abundance,\footnote{Throughout this paper, we adopt the notation
[X/Y] to represent the relative number density of elements $X$ and $Y$
on a logarithmic and solar abundance scale. Explicitly,
[X/Y]~$=~\log_{10}(N({\rm X})/N({\rm Y}))-\log_{10}(n({\rm X})/n({\rm Y}))_{\odot}$.}
since \OI\ accurately traces the \HI\ gas due to charge transfer reactions \citep{FieSte71}.
Furthermore, we only consider the total column density of Components 3, 4 and 5
(the only components where \OI\ is detected).
Using a solar oxygen abundance of log$_{10}$\,(O/H)$_{\odot}$ + 12 = 8.69 \citep{Asp09}, we estimate
an oxygen abundance ${\rm [O/H]}=-2.769\pm0.028$, which compares well
to that reported by \citet{Kir03}, of ${\rm [O/H]}=-2.79\pm0.05$.

\begin{figure*}
  \centering
 {\includegraphics[angle=0,width=160mm]{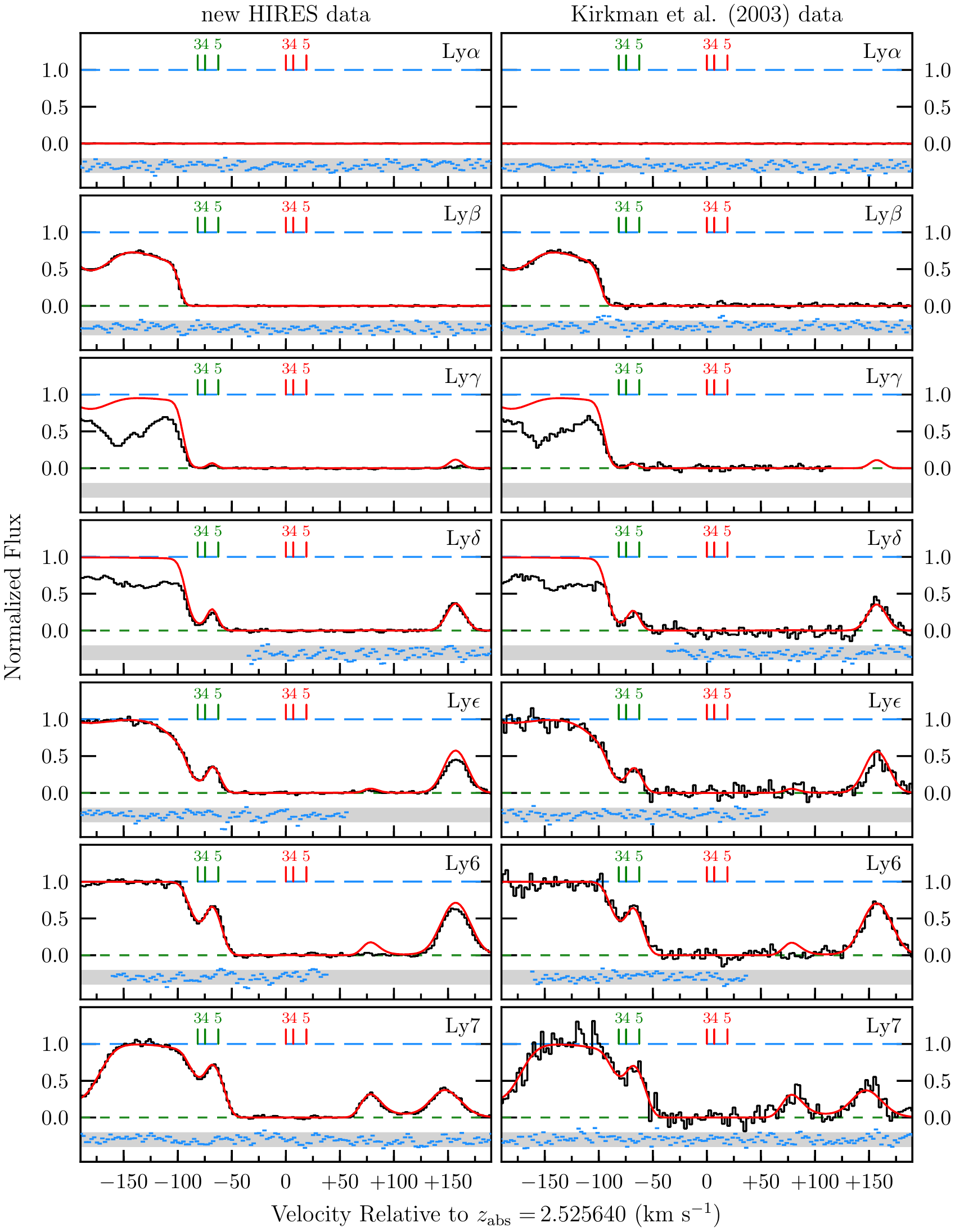}}\\
  \caption{
The Lyman series lines of the absorption system at \zabs\ towards the QSO Q1243$+$307 are
shown (black histograms), overlaid with the best-fitting model profile (red lines). The left panels
display the newly acquired Keck HIRES data, which can be compared with the \citet{Kir03}
old Keck HIRES data shown in the right panels. In all panels, the data and models are normalized
to the best-fitting continuum (long dashed blue lines), and the fitted zero level has been removed
(short dashed green lines). The red and green tick marks above the spectrum indicate the locations
of the three primary absorption components seen in \HI\ and \DI, respectively
(denoted Components 3, 4, and 5 in Table~\ref{tab:compstruct}, as indicated above each tick mark). Although only three tick marks are
shown, we note that the model profile presented in each panel (i.e. the red curve) includes all model
components listed in Table~\ref{tab:compstruct}.
The blue points below each spectrum are the normalized fit residuals, (data--model)/error, of all pixels used in the analysis, and the gray band represents a confidence interval of $\pm2\sigma$.
A label in the top right corner of every panel indicates the Lyman series transition shown.
  }
  \label{fig:lyseriesa}
\end{figure*}

\begin{figure*}
  \centering
 {\includegraphics[angle=0,width=160mm]{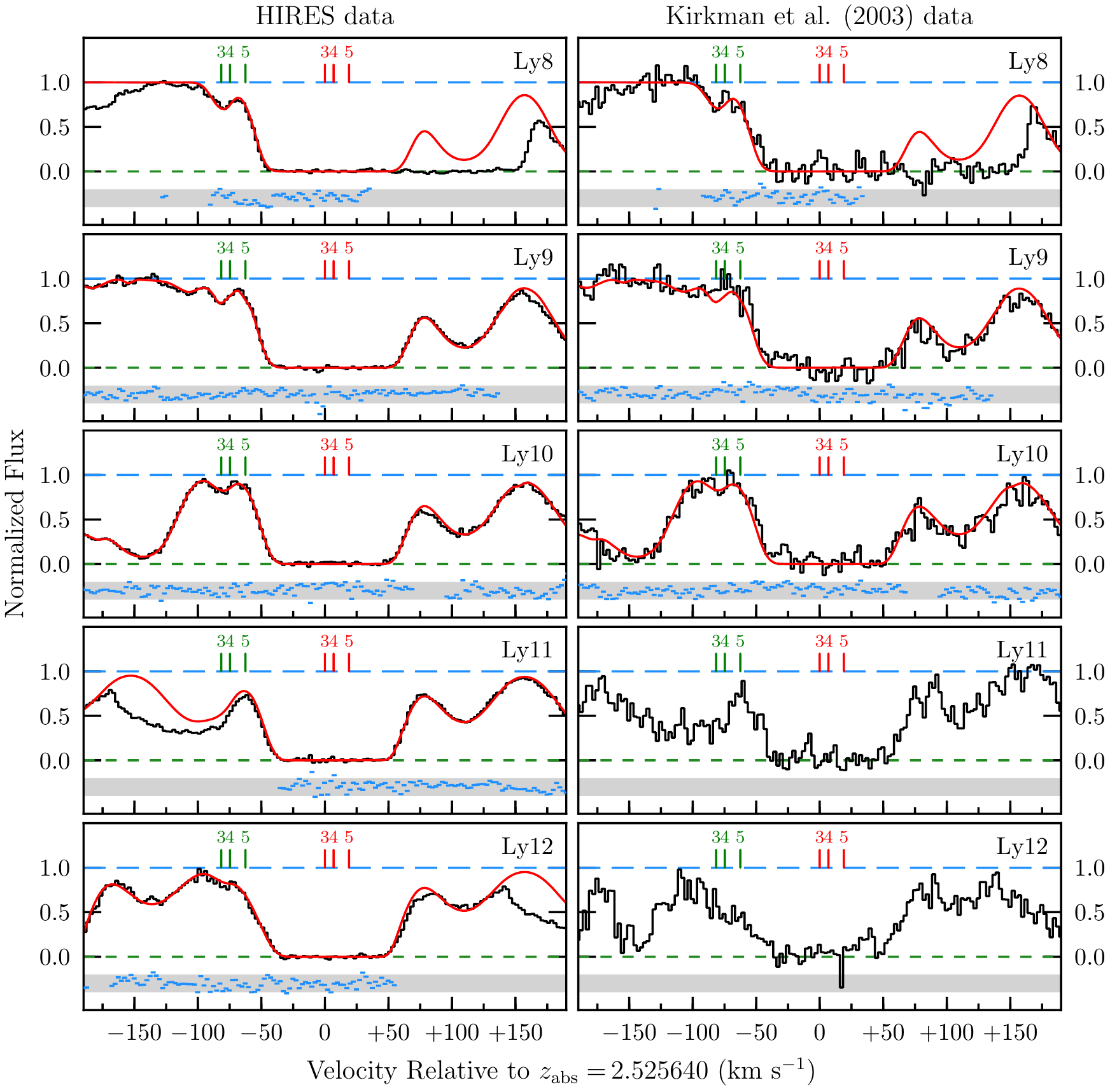}}\\
  \caption{
Same as Figure~\ref{fig:lyseriesa}, for the higher order Lyman series lines. Both Ly11 and Ly12 from the
\citet{Kir03} data are not used in our analysis, but the data are shown here for reference without
a model overplotted.
We note that the \citet{Kir03} data consist of a total exposure time of 55,800~s, while the
new data we report here were obtained with a total exposure time of 13,800~s.
Even though the new data were taken with a quarter of the exposure time of the old data, the new data
have considerably higher signal-to-noise ratio, especially near the highest order Lyman series
lines (corresponding to the weakest \DI\ absorption lines) near an observed wavelength
$\lambda_{\rm obs} = 3230$\,\AA. This comparison offers a clear
demonstration of the increased sensitivity made possible by the HIRES detector upgrade.
  }
  \label{fig:lyseriesb}
\end{figure*}

Like our previous analyses, we force all \HI\ components to share the same $N$(\DI)/$N$(\HI) ratio.
The initial starting parameter value of the logarithmic $N$(\DI)/$N$(\HI) ratio is drawn from a uniform
distribution over the range ($-4.7$, $-4.5$). Our analysis uses a total of eight \DI\ Lyman
series lines, including \Lyb, Ly$\epsilon$--Ly10, and Ly12.\footnote{We do not include the
Ly12 transition from the \citet{Kir03} data in our analysis, since the data are noisy and the
blue wing shows a different structure to the new HIRES data.} Since our data near the Lyman Limit
is of considerably higher signal-to-noise ratio than the data obtained by
\citet[][see Figures~\ref{fig:lyseriesa} and \ref{fig:lyseriesb}]{Kir03}, we have
identified several unrelated absorption line systems that are blended
by chance with three of the \DI\ absorption lines (Ly$\epsilon$, Ly7, and Ly9;
see Appendix~\ref{sec:appendix}); only one of these blends was discernible and accounted for
in the lower signal-to-noise data presented by \citet{Kir03}. This highlights the importance of
obtaining high signal-to-noise data down to the Lyman Limit which, in this case, corresponds to
an observed wavelength of $\sim3215$~\AA. We have
fully accounted for these blends by fitting the associated line profiles of each system, the details
of which are provided in Appendix~\ref{sec:appendix}. We present the best-fit model to the Lyman
series lines in Figures~\ref{fig:lyseriesa} and \ref{fig:lyseriesb}.

Throughout the analysis described above, we adopt a blind analysis strategy,
whereby the logarithmic $N$(\DI)/$N$(\HI) ratio is not revealed until after the model fitting is complete.
To ensure that the global minimum $\chi$-squared has been reached,
we perform 2000 Monte Carlo simulations, where the starting parameters of each simulation correspond
to the model fitting parameters perturbed by twice the covariance matrix. We also redraw
a new logarithmic $N$(\DI)/$N$(\HI) ratio from a uniform distribution over the range ($-4.7$, $-4.5$) for
each simulation. This process ensures that no memory of the starting parameters affects the
final result. We then identify the realization that gives the minimum $\chi$-squared, unblind the
logarithmic $N$(\DI)/$N$(\HI) ratio and refer to this as the `best-fitting' model throughout our analysis.
We present the best-fitting model absorption profile parameters
in Table~\ref{tab:compstruct}. The resulting $\chi$-squared/dof\footnote{\label{foot:chisq}We note that our analysis does not account for correlations
between neighboring spectral pixels, and therefore the reported $\chi$-squared value
is likely underestimated, as discussed previously by \citet{Coo14,Coo16}.} of this model is
$12115/13697\simeq0.885$.

\subsection{Sensitivity of D/H to the \Lya\ continuum}
\label{sec:senslya}

After the analysis was complete, we tested
how the value and precision of D/H was affected by the adopted fitting range
around \Lya. We increased the fitted range to include all pixels and blends
within $\pm1250~{\rm km~s^{-1}}$ relative to the sub-DLAs redshift (i.e. all
pixels with a residual intensity less than 90 per cent of the quasar continuum).
The central value of D/H was unchanged and the error was increased by just
$\sim2.7$ per cent, confirming that the final precision of D/H is
relatively insensitive to the pixels outside of the original velocity
window (i.e. beyond $-470\le v/{\rm km~s^{-1}}\le+580$).

\subsection{Sensitivity of D/H to the component structure}
\label{sec:sensmetals}

Subsequent to unblinding, we appreciated
that some of the `satellite' absorption components of the \CII\,$\lambda1334$ and
\SiII\,$\lambda1260$ absorption lines (in particular, Components 1 and 6) were
underfit. We therefore included additional absorption components in the metal
lines to test if the value or error of the \DI/\HI\ value reported here was affected.
By including an additional four absorption components
(at velocities $-38.8$, $-5.8$, $+25.2$, and $45.7$ km~s$^{-1}$ relative to Component 3),
we found that the resulting chi-squared was significantly improved (11972/13678). The
central D/H value of this model did not change, and the uncertainty increased by
6.4 per cent (from $\pm0.015$ to $\pm0.016$ in the log; see Equation~\ref{eqn:dihi}).
This is a negligible increase, and is not reflected in the reported error budget on \DI/\HI.

\section{The Precision Sample}
\label{sec:sample}

\subsection{Sample Definition}

As outlined by \citet{Coo14}, an absorption line system that
meets the following selection criteria is almost ideal for
obtaining a high precision measurement of the deuterium abundance.
Specifically, all D/H measures considered in this paper have:
(1) An \HI\ column density in excess of $10^{19}~{\rm cm}^{-2}$. In this column
density regime, the \Lya\ absorption line is damped by the Lorentzian term
of the Voigt profile and a unique
value of \NHI\ can be determined from the wings of the line profile;
(2) a \Lya\ profile that is not severely blended by
contaminating absorption;
(3) at least two unblended and optically thin \DI\ transitions from which the total
column density of neutral deuterium can be determined;
(4) data that were acquired with a high resolution echelle spectrograph, and
of high signal-to-noise ratio ($>10$~pixel$^{-1}$) near both \Lya\
and the weakest \DI\ absorption line used in the analysis.
We further add that all systems have been self-consistently
analyzed by our group; including other recent determinations
\citep{Bal16,Rie17,Zav17} may introduce unaccounted
for systematic errors because of the different analysis
techniques adopted.

\subsection{A new deuterium abundance measurement}

The absorption system at \zabs\ towards the quasar Q1243$+$307 satisfies all of the
above criteria, and we now include this new measurement to the \textit{Precision Sample}
of D/H measures.
Using the analysis procedure described in Section~\ref{sec:analysis},
we determine the logarithmic ratio of neutral deuterium to neutral hydrogen
atoms of this absorption system to be:
\begin{equation}
\label{eqn:dihi}
\log_{10}\,N({\rm D\,\textsc{i}})/N({\rm H\,\textsc{i}}) = -4.622\pm0.015
\end{equation}
This number compares very well with the central value reported by \citet{Kir03},
$\log_{10}\,N({\rm D\,\textsc{i}})/N({\rm H\,\textsc{i}}) = -4.617^{+0.058}_{-0.048}$.
The factor of $3.5$ improvement on the $N$(\DI)/$N$(\HI) measurement precision
that we report here is largely due to the much higher signal-to-noise ratio of the new data
near the highest order Lyman series lines, compared to the \citet{Kir03} data.
It is reassuring that the central values reported by both analyses are mutually
consistent.

The $N$(\DI)/$N$(\HI) value found here is consistent with the previous
measures by \citet{Coo14,Coo16}, which are all collected in Table~\ref{tab:dhmeasures}. This new
$N$(\DI)/$N$(\HI) value comes from one of the lowest metallicity systems currently known,
making it a key measurement to assess whether or not $N$(\DI)/$N$(\HI) varies with metallicity.
We note that the estimated $N$(\DI)/$N$(\HI) value of this system is the lowest of the seven
systems analyzed by our group, and is derived from one of the lowest \HI\ column
density absorbers that we have considered so far.
Despite the relatively low \HI\ column density, we emphasize that the expected
ionization correction is $<0.001$~dex when the \HI\ column
density is $10^{19.76}~{\rm atoms~cm}^{-2}$ \citep{CooPet16}.

\begin{figure*}
  \centering
 {\includegraphics[angle=0,width=87mm]{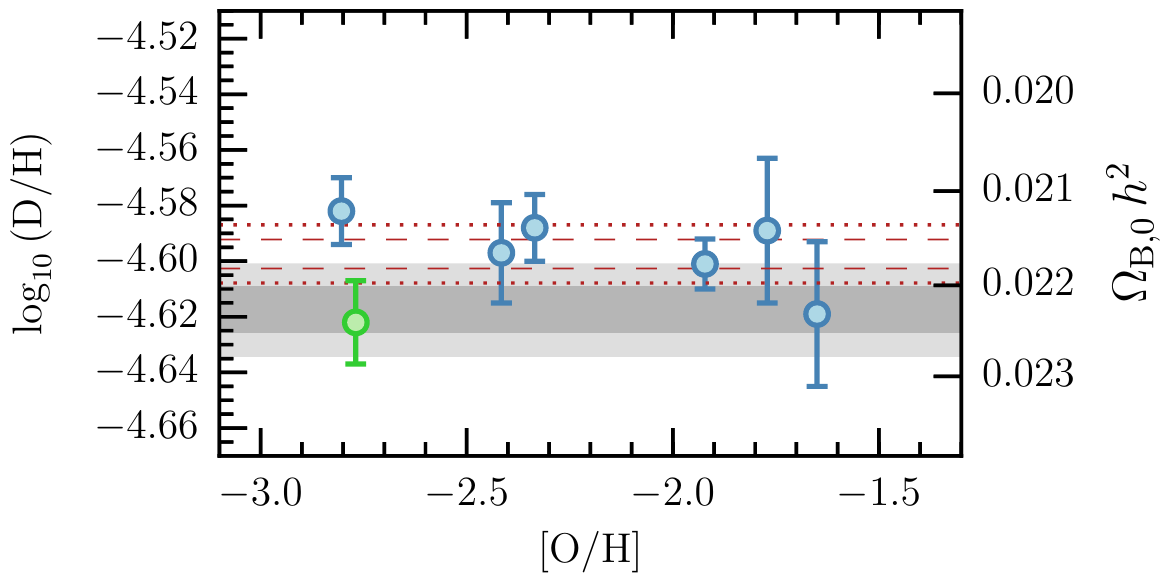}
 \hspace{6mm}\includegraphics[angle=0,width=87mm]{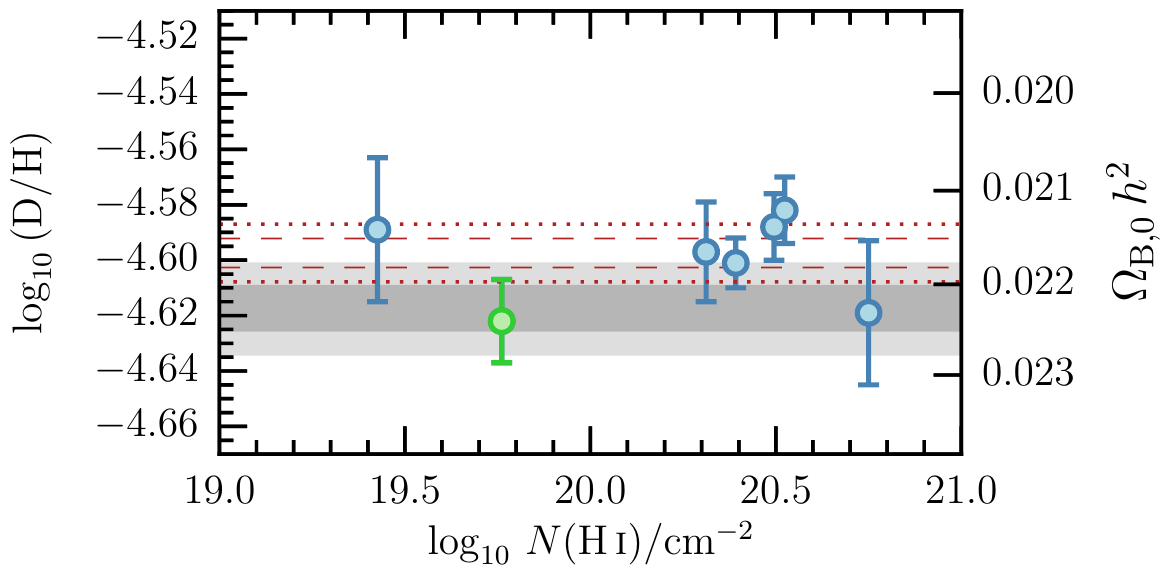}}\\
  \caption{Our sample of seven high precision D/H measures is shown (symbols with error bars); the green symbol represents the new measure that we report here. The weighted mean value of these seven measures is shown by the red dashed and dotted lines, which represent the 68 and 95 per cent confidence levels, respectively. The left and right panels show the dependence of D/H on the oxygen abundance and neutral hydrogen column density, respectively. Assuming the Standard Model of cosmology and particle physics, the right vertical axis of each panel shows the conversion from D/H to the universal baryon density. This conversion uses the \citet{Mar16} theoretical determination of the \dpg\ cross-section. The dark and light shaded bands correspond to the 68 and 95 per cent confidence bounds on the baryon density derived from the CMB \citep{Efs15}.}
  \label{fig:measures}
\end{figure*}

The Precision Sample of D/H measurements is shown in the left
and right panels of Figure~\ref{fig:measures} as a function of metallicity and \HI\
column density, respectively.
We first draw attention to the subtle decrease
of D/H with increasing metallicity suggested by the six blue symbols in
Figure~\ref{fig:measures} (see also, \citealt{Coo16}). The new measurement that we report here (indicated by
the green symbol in Figure~\ref{fig:measures}) does not support this trend.
There is also no apparent trend of D/H with \HI\ column density. In what follows,
we therefore assume that $N$(\DI)/$N$(\HI)$~\equiv~$D\,/\,H.

\begin{table*}
\begin{center}
    \caption{\textsc{precision d/h measures considered in this paper}}
    \hspace{-0.6cm}\begin{tabular}{@{}lccccc}
    \hline
   \multicolumn{1}{c}{QSO}
& \multicolumn{1}{c}{$z_{\rm em}$}
& \multicolumn{1}{c}{$z_{\rm abs}$}
& \multicolumn{1}{c}{log$_{10}$~$N$(\HI)/cm$^{-2}$}
& \multicolumn{1}{c}{[O/H]$^{\rm a}$}
& \multicolumn{1}{c}{log$_{10}$~$N$(\DI)/$N$(\HI)}\\
  \hline
HS\,0105$+$1619         &  $2.652$  &  $2.53651$  &  $19.426\pm0.006$  &  $-1.771\pm0.021$  &  $-4.589\pm0.026$  \\
Q0913$+$072               &  $2.785$  &  $2.61829$  &  $20.312\pm0.008$  &  $-2.416\pm0.011$  &  $-4.597\pm0.018$  \\
Q1243$+$307               &  $2.558$  &  $2.52564$  &  $19.761\pm0.026$  &  $-2.769\pm0.028$  &  $-4.622\pm0.015$  \\
SDSS~J1358$+$0349  &  $2.894$  &  $2.85305$  &  $20.524\pm0.006$   &  $-2.804\pm0.015$  &  $-4.582\pm0.012$  \\
SDSS~J1358$+$6522  &  $3.173$  &  $3.06726$  &  $20.495\pm0.008$   &  $-2.335\pm0.022$  &  $-4.588\pm0.012$  \\
SDSS~J1419$+$0829  &  $3.030$  &  $3.04973$  &  $20.392\pm0.003$   &  $-1.922\pm0.010$  &  $-4.601\pm0.009$  \\
SDSS~J1558$-$0031   &  $2.823$  &  $2.70242$  &  $20.75\pm0.03$       &  $-1.650\pm0.040$  &  $-4.619\pm0.026$  \\
  \hline
    \end{tabular}
    \label{tab:dhmeasures}

$^{\rm a}${We adopt the solar value log$_{10}$\,(O/H) + 12 = 8.69 \citep{Asp09}.}\\
\end{center}
\end{table*}

\subsection{Intrinsic Scatter}

Even though the seven measurements considered here show no apparent
trend with metallicity or \HI\ column density, there may still be an intrinsic
scatter of these D/H measurements that could be due to systematics that
are currently unaccounted for. Such an `excess' dispersion in D/H abundance
measurements was originally noted by \citet{Ste01} for an earlier, and
more heterogeneous, sample of
D/H values. Indeed, a simple $\chi^{2}$ test reveals that these seven measures
are statistically consistent (i.e. within $2\sigma$) of being drawn from a
constant D/H value. This suggests that the intrinsic scatter among the
measurements must be low, and we now explore this in further detail.

Suppose that each measured D/H value, $d_{i}$, with uncertainty $\sigma_{i}$
has a corresponding `true' value, $d_{\rm T}$. The probability that a given observation
arises from the true value is given by

\begin{equation}
{\rm Pr}(d_{i}|d_{\rm T}) = \frac{1}{\sqrt{2\pi}\sigma_{i}}\exp\bigg(-\frac{(d_{i}-d_{\rm T})^2}{2\sigma_{i}^2}\bigg)
\end{equation}

Similarly, if the true values are drawn from an `intrinsic' distribution with central value $DH_{\rm P}$
and scatter $\sigma$, the probability that a true value is drawn from the intrinsic distribution is

\begin{equation}
{\rm Pr}(d_{\rm T}|DH_{\rm P}) = \frac{1}{\sqrt{2\pi}\sigma}\exp\bigg(-\frac{(d_{\rm T}-DH_{\rm P})^2}{2\sigma^2}\bigg)
\end{equation}

Therefore, the probability of obtaining a measured D/H value, $d_{i}$, given our intrinsic model
is found by integrating over all possible true values

\begin{eqnarray}
{\rm Pr}(d_{i}|DH_{\rm P}) &=& \int_{-\infty}^{+\infty} {\rm Pr}(d_{i}|d_{\rm T})~{\rm Pr}(d_{\rm T}|DH_{\rm P})~{\rm d}d_{\rm T} \nonumber \\
{\rm Pr}(d_{i}|DH_{\rm P}) &=& \frac{1}{\sqrt{2\pi(\sigma_{i}^2+\sigma^2)}}\exp\bigg(-\frac{(d_{i} - DH_{\rm P})^2}{2(\sigma_{i}^2+\sigma^2)}\bigg)
\end{eqnarray}
and the log-likelihood function is then given by
\begin{equation}
\label{eqn:likelihood}
{\cal L} = \log \bigg[\prod_{i}~{\rm Pr}(d_{i}|DH_{\rm P})\bigg]
\end{equation}
Using a brute force method, we solve for the parameter values ($DH_{\rm P}$ and $\sigma$)
that maximize the likelihood function in Equation~\ref{eqn:likelihood}, based on the seven measures
listed in Table~\ref{tab:dhmeasures}. The maximum likelihood parameter values are
\begin{eqnarray}
DH_{\rm P}~&=&~-4.5976\pm0.0072 \\
\sigma~&\le&~0.027\qquad({\rm 95\%\ confidence})
\end{eqnarray}
Note that the intrinsic dispersion, $\sigma$, has a maximum likelihood value
of zero; we therefore quote a $2\sigma$ upper limit. The above likelihood
analysis indicates that there is very little intrinsic scatter in our defined sample of
consistently analyzed D/H measures. We therefore speculate that the
original excess scatter noted by \citet{Ste01} is probably due to a
combination of the different analysis techniques employed by different authors
and the use of absorption line systems that were not well-suited for measuring D/H.
Together, these factors probably resulted in underestimates of the true errors
in the values of D/H reported.

\section{Cosmological Consequences}
\label{sec:cosmology}

\subsection{The primordial deuterium abundance}

Based on the analysis above, we conclude that the
seven D/H measurements considered here are drawn from the
same value, and a weighted mean of these measures gives
our best estimate of the primordial deuterium abundance:\footnote{These values
and their errors are unaffected by the small error increases resulting from the changes
to our fitting procedure, as discussed in Sections~\ref{sec:senslya} and \ref{sec:sensmetals}).}
\begin{equation}
\label{eqn:dhp}
\log_{10}\,({\rm D/H})_{\rm P} = -4.5974\pm0.0052
\end{equation}
or, expressed as a linear quantity:
\begin{equation}
10^{5}\,({\rm D/H})_{\rm P} = 2.527\pm0.030
\end{equation}
This value corresponds to a $\sim1$ per cent determination of the
primordial deuterium abundance, and is shown in Figure~\ref{fig:measures}
by the dashed and dotted horizontal lines to represent the 68 and 95 per cent
confidence regions, respectively. Our determination of the primordial deuterium
abundance quoted here has not changed much from our previous estimate in
\citet{Coo16}; as discussed above, the new value is in mutual agreement with
the previous six measures and is of comparable precision. We therefore conclude
that the primordial deuterium abundance quoted here is robust.

\subsection{Testing the Standard Model}

In order to compare this measurement to the latest \textit{Planck}
CMB results, we must first convert our estimate of \dhp\ to the
baryon-to-photon ratio, $\eta$. To do this, we use the BBN calculations
described by \citet[][see also, \citealt{NolBur00,NolHol11}]{Coo16}, assuming
the \citet{Mar16} \dpg\ reaction rate. For
the case of the Standard Model, we deduce a baryon-to-photon ratio of
\begin{equation}
10^{10}\,\eta\equiv\eta_{10}=5.931\pm0.051
\end{equation}
which includes the uncertainty of the nuclear data that are used as input
to the BBN calculations.

We can now convert this value of the baryon-to-photon ratio into an
estimate of the cosmic density of baryons using the formula
$\eta_{10}=(273.78\pm0.18)\times\Omega_{\rm B,0}\,h^{2}$
\citep{Ste06} which, for the Standard Model, gives the value:
\begin{equation}
\label{eqn:obhhbbn}
100\,\Omega_{\rm B,0}\,h^{2}({\rm BBN}) = 2.166\pm0.015\pm0.011
\end{equation}
where the first error term includes the uncertainty in the
measurement of (D/H)$_{\rm P}$, and the second error term
provides the uncertainty in the BBN calculations.

\begin{figure}
  \centering
 {\includegraphics[angle=0,width=80mm]{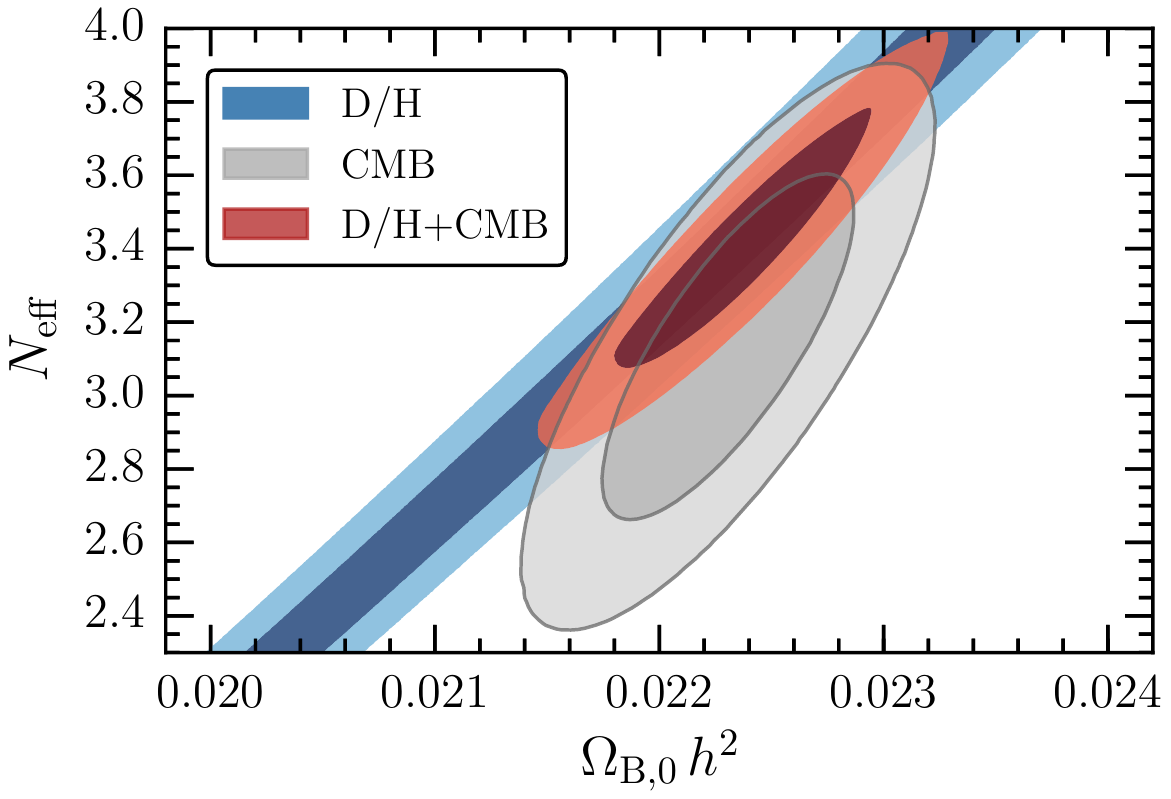}}\\
  \caption{
Comparing the expansion rate (parameterized by \neff) and the
cosmic density of baryons (\obhh) from BBN (blue contours) and
CMB (gray contours). The red contours indicate the combined
confidence bounds. The dark and light shades
illustrate the 68\%\ and 95\%\ confidence contours, respectively.
  }
  \label{fig:neffobhh}
\end{figure}

The BBN inferred value of the cosmic baryon density is somewhat lower than the CMB value \citep[][see gray bands in Figure~\ref{fig:measures}]{Efs15}.\footnote{The quoted value of \obhh(CMB) corresponds to the \textit{Planck} TT+lowP+lensing analysis, listed in the second data column of Table 4 from \citet{Efs15}. As demonstrated by these authors, the CMB value of \obhh\ is robust to simple extensions of the base $\Lambda$CDM model.}
\begin{equation}
100\,\Omega_{\rm B,0}\,h^{2}({\rm CMB}) = 2.226\pm0.023
\end{equation}
This difference corresponds to a $2\sigma$ discrepancy. As discussed by \citet{Coo16},
there is still some tension between the computed and the empirically measured \dpg\ cross-section
that is used as input into the BBN calculations. Adopting the empirically measured \dpg\
cross-section proposed by \citet{Ade11}, we estimate a cosmic baryon abundance of
\begin{equation}
100\,\Omega_{\rm B,0}\,h^{2}({\rm BBN}) = 2.235\pm0.016\pm0.033
\end{equation}
where the error terms have the same meaning as Equation~\ref{eqn:obhhbbn}.
This value is in better agreement with the CMB measurement. In the near future,
we expect to hear the results of a \dpg\ measurement campaign from the
Laboratory for Underground Nuclear Astrophysics (LUNA; \citealt{Gus14,Koc16}),
which will provide new insight to this important BBN reaction rate.

An alternative possibility to bring closer together the BBN and CMB determinations
of \obhh\ is to consider simple
extensions to the Standard Model, such as a change in the expansion rate,
parameterized by the effective number of neutrino species, \neff.
In Figure~\ref{fig:neffobhh}, we present the confidence contours of the
baryon density and the effective number of neutrino families for the
\dhp\ estimate that we report here (blue band)\footnote{Assuming the
\citet{Mar16} \dpg\ reaction rate.} in addition to the
\textit{Planck} CMB results (gray ellipse). The combined confidence
contours are displayed in red, where the central value and uncertainty
of these parameters are (95 per cent confidence limits):
\begin{eqnarray}
100\,\Omega_{\rm B,0}\,h^{2} &=& 2.237\pm0.070\\
N_{\rm eff} &=& 3.41\pm0.45
\end{eqnarray}
where the Standard Model value of the effective number of neutrino
species is \neff=3.046 \citep[][see also, \citealt{Gro15}]{Man05}.
We therefore conclude that our results are consistent (within $2\sigma$)
with the Standard Model of cosmology and particle physics.

\section{Summary and Conclusions}
\label{sec:conc}

We have presented a reanalysis of the \zabs\ absorption system along the
line-of-sight to Q1243$+$307 using data previously reported by \citet{Kir03},
combined with new data acquired with the Keck HIRES echelle spectrograph,
to deduce the deuterium abundance of an extremely metal-poor gas cloud.
Our conclusions are summarized as follows:\\

\noindent ~~(i) Using the upgraded, ultraviolet sensitive detector on the HIRES echelle
spectrograph, we have obtained exquisite, high signal-to-noise data down to the Lyman
limit of Q1243+307 near the observed wavelength $\lambda_{\rm obs} = 3215$\,\AA.
Combined with archival data, we estimate the oxygen abundance of this absorber to be
[O/H]=$-2.769\pm0.028$, which is among the lowest
metallicity environment where the deuterium abundance has been measured.

\smallskip

\noindent ~~(ii) On the basis of eight \DI\ Lyman series absorption
lines, we infer that the deuterium abundance of this system is
$\log_{10}\,N({\rm D\,\textsc{i}})/N({\rm H\,\textsc{i}})~= -4.622\pm0.015$,
which is in excellent agreement with the value reported by
\citet[][$\log_{10}\,N({\rm D\,\textsc{i}})/N({\rm H\,\textsc{i}})~= -4.617^{+0.058}_{-0.048}$]{Kir03}.
Our measure has therefore improved the precision of this one measurement by a factor of $\sim3.5$.

\smallskip

\noindent ~~(iii) Combining our new measurement with the six homogeneously
analyzed systems previously reported by our group, we use a maximum likelihood
technique to determine the intrinsic dispersion of our D/H sample. We find that the
seven D/H measures are consistent with being drawn from a constant value
(i.e. no intrinsic dispersion).
Thus, the excess dispersion of D/H values previously recognized by \citet{Ste01}
can be attributed to the different analysis techniques adopted by different
authors, and the use of absorption line systems that were not well-suited for
precisely measuring the primordial abundance of deuterium.
Together, these factors probably resulted in underestimates of
the true errors in the values of D/H reported.

\smallskip

\noindent ~~(iv) We also find that these seven D/H values do not correlate
with [O/H] or \NHI, strongly suggesting that
our sample of D/H measures corresponds to the
primordial abundance of deuterium.

\smallskip

\noindent ~~(v) Based on the seven systems analyzed by our group,
we estimate that the primordial deuterium abundance is
$\log_{10}\,({\rm D/H})_{\rm P} = -4.5974\pm0.0052$,
or, expressed as a linear quantity
$10^{5}\,({\rm D/H})_{\rm P} = 2.527\pm0.030$.
Thanks to modern instrumentation, and a careful analysis, it is now possible to
pin down the primordial abundance of deuterium with $\sim\!\!1$ percent precision,
using just seven D/H measures.

\smallskip

\noindent ~~(vi) Using a suite of BBN calculations that use the latest
nuclear physics input (previously described by \citealt{Coo16}),
we estimate that the cosmic abundance of baryons is
$100\,\Omega_{\rm B,0}\,h^{2}({\rm BBN}) = 2.166\pm0.015\pm0.011$, where
we separately quote the error associated with the measurement (former) and
the BBN calculation (latter). This value is based on the \dpg\ reaction rate
computed by \citet{Mar16}, and differs from the \textit{Planck} CMB value
by $2\sigma$. Alternatively, using an empirically determined \dpg\ reaction rate,
we estimate a baryon density of
$100\,\Omega_{\rm B,0}\,h^{2}({\rm BBN}) = 2.235\pm0.016\pm0.033$,
which is in better agreement with the CMB, albeit with larger errors.

\smallskip

\noindent ~~(vii) We also perform a joint analysis of D/H and the \textit{Planck} CMB
data to place a bound on the effective number of neutrino species. Our joint constraints
on the cosmic baryon abundance and effective number of neutrino families are
$100\,\Omega_{\rm B,0}\,h^{2} = 2.237\pm0.070$ and
$N_{\rm eff} = 3.41\pm0.45$, respectively ($95\%$ confidence).

\smallskip


Given that the CMB is now limited by cosmic variance at scales $l\lesssim10^{3}$ --
the multipole regime where the temperature fluctuations are very sensitive
to the baryon density -- it will become increasingly difficult to \textit{significantly}
improve the precision of \obhh\ derived from the CMB. Based on just seven
determinations of the deuterium abundance of near-pristine gas clouds,
we have reached a one percent precision on the primordial deuterium
abundance, corresponding to a sub-percent level precision on \obhh;
this level of precision is comparable to, or somewhat better than, that
reached by the latest CMB constraints.
In addition, there are exciting opportunities in the immediate future to further increase
the statistics of D/H with the The Echelle SPectrograph for Rocky Exoplanet
and Stable Spectroscopic Observations (ESPRESSO) spectrograph on the European Southern Observatory Very
Large Telescope, and potentially in the longer term with the 30-40\,m class telescopes.

\acknowledgments

We are grateful to the staff astronomers at Keck  
Observatory for their assistance with the observations, and
to Jason X. Prochaska, Tom Barlow, and Michael T. Murphy for providing some
of the software that was used to reduce the data.
We also thank an anonymous referee who provided helpful
suggestions that improved the presentation of this work, following
a referee who was unable to respond in a timely manner.
During this work, R.~J.~C. was supported by a Royal Society
University Research Fellowship.
R.~J.~C. acknowledges support from STFC (ST/L00075X/1, ST/P000541/1).
C.~C.~S. has been supported by grant AST-1313472 from the U.S. NSF.
This research was also supported by a NASA Keck PI Data Award,
administered by the NASA Exoplanet Science Institute.
Data presented herein were obtained at the W. M. Keck Observatory
from telescope time partially allocated to the National Aeronautics and Space
Administration through the agency's scientific partnership with the
California Institute of Technology and the University of California.
The Observatory was made possible by the generous financial
support of the W. M. Keck Foundation. We thank the Hawaiian
people for the opportunity to observe from Mauna Kea;
without their hospitality, this work would not have been possible.
This work used the DiRAC Data Centric system at Durham University,
operated by the Institute for Computational Cosmology on behalf of the
STFC DiRAC HPC Facility (www.dirac.ac.uk). This equipment was funded
by BIS National E-infrastructure capital grant ST/K00042X/1, STFC capital
grant ST/H008519/1, and STFC DiRAC Operations grant ST/K003267/1
and Durham University. DiRAC is part of the National E-Infrastructure.
This research has made use of NASA's Astrophysics Data System.
 R.J.C. thanks S.O.C. for her impeccable timing and
invaluable insight.

%

\vspace{5mm}
\facility{Keck(HIRES)}


\software{ Astropy \citep{Ast13},
		Matplotlib \citep{Hun07},
		NumPy \citep{van11}
        		}



\appendix
\section{Accounting for line blending of the D {\sc i} Lyman series profiles}
\label{sec:appendix}

In this appendix, we present the model profiles of several unrelated absorption systems that are coincident (by chance) with the Lyman series absorption lines of the sub-DLA at \zabs. To identify these blended absorption systems, we systematically cross-checked every line of the sub-DLA's \DI\ Lyman series, to identify potential contamination with the Lyman series of an unrelated, higher redshift \HI\ system. We also cross-checked each line of the sub-DLA's \DI\ Lyman series against the expected positions of the metal-strong absorption line system located at $z_{\rm abs}=2.053206$.
In all cases, we use unblended lines of the contaminant system to determine more accurately the shape and depth of the absorption line that is blended with the sub-DLA. We have identified three \DI\ Lyman series lines that are affected by blending; these include Ly$\epsilon$, Ly7, and Ly9, which we now discuss in turn.

\begin{figure}[!htb]
    \centering
    \begin{minipage}{0.46\textwidth}
        \centering
        \vspace{1.1cm}
        \includegraphics[width=\linewidth]{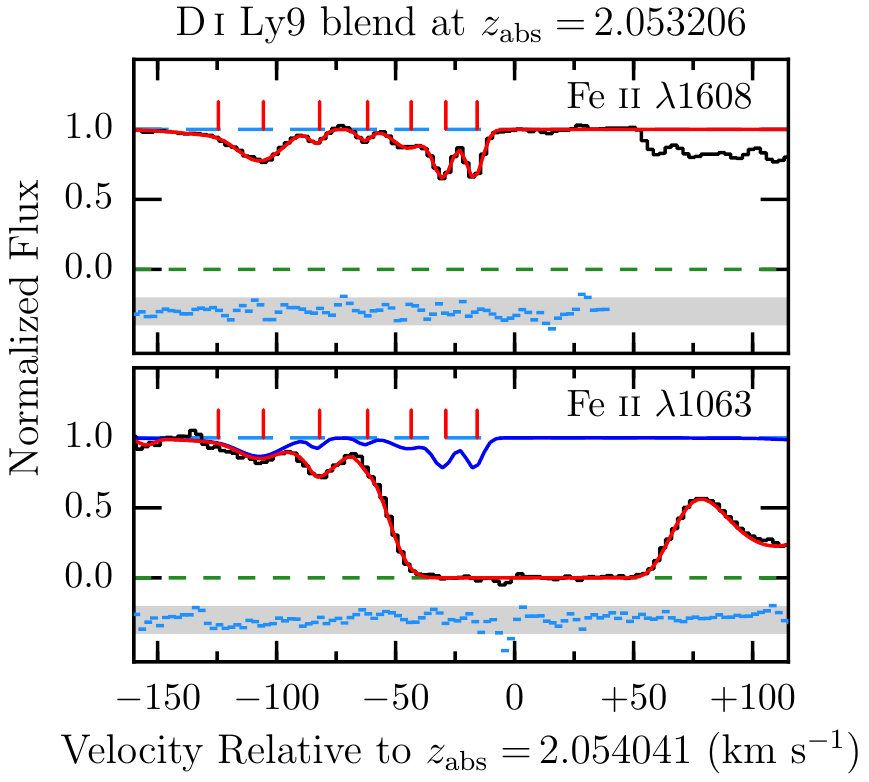}
        \caption{\DI\ Ly9 of the sub-DLA (bottom panel) is blended with an \FeII\,$\lambda1063$ complex at $z_{\rm abs}=2.053206$. The top panel shows the corresponding \FeII\,$\lambda1608$ absorption line of the contaminant system. In both panels, the red line shows the best model fit to the data (black histogram). The blue curve in the bottom panel shows the contribution of the blend to the total absorption profile. The long blue dashed lines represent the continuum levels, while the short green dashed line indicates the zero levels. The red tick marks above each spectrum indicate the absorption components of the blend. The blue points below each spectrum are the normalized fit residuals, (data--model)/error, of all pixels used in the analysis, and the gray band represents a confidence interval of $\pm2\sigma$.}
        \label{fig:blend1}
    \end{minipage}%
    \hspace{0.03\textwidth}
    \begin{minipage}{0.46\textwidth}
        \centering
        \includegraphics[width=\linewidth]{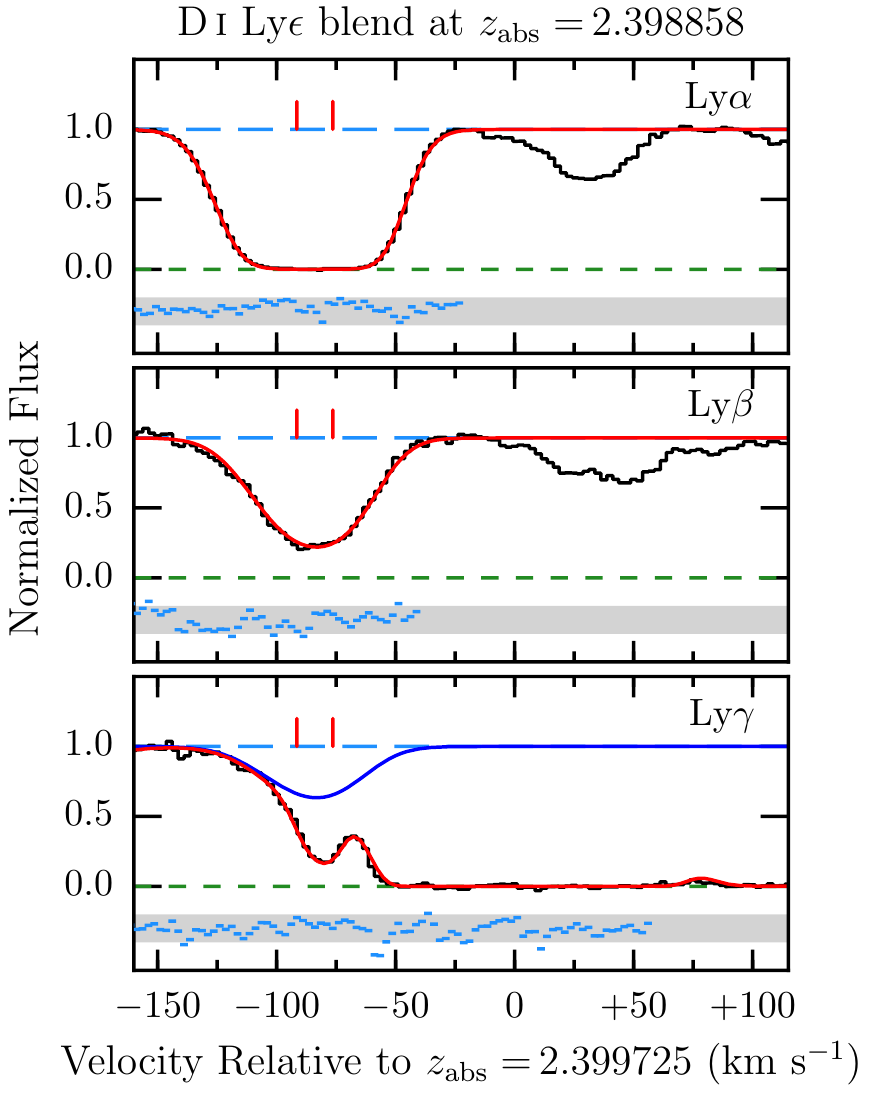}
        \caption{Same as Figure~\ref{fig:blend1}, but illustrating the blended \DI\ Ly$\epsilon$ transition of the sub-DLA (bottom panel), due to an \HI\ Ly$\gamma$ absorption line of an unrelated system at $z_{\rm abs}=2.3988$ (see blue profile in the bottom panel). The corresponding Ly$\alpha$ and Ly$\beta$ lines of this blend are shown in the top and middle panel, respectively.}
        \label{fig:blend2}
    \end{minipage}
\end{figure}

In Figure~\ref{fig:blend1}, we show the \DI\ Ly9 blend, which is due to \FeII\,$\lambda1063$\,\AA\ absorption from an unrelated absorption complex at $z_{\rm abs}=2.053206$. The \FeII\ absorption is well modeled by the \FeII\,$\lambda1608$\,\AA\ absorption, which falls in a clean part of the spectrum outside of the \Lya\ forest and has roughly the same strength as the \FeII\,$\lambda1063$\,\AA\ absorption.

The \DI\ Ly$\epsilon$ blend is shown in Figure~\ref{fig:blend2},
and comprises two components at
$z_{\rm abs}=2.39886\pm0.00005$ and
$z_{\rm abs}=2.39869\pm0.00013$, with \HI\ column densities of
log$_{10}$~\NHI/cm$^{-2} = 14.22\pm0.43$ and
log$_{10}$~\NHI/cm$^{-2} = 14.18\pm0.48$, respectively.
The total Doppler parameters of these absorption lines are
$b_{\rm tot}=21.4\pm1.5~{\rm km~s}^{-1}$ and
$b_{\rm tot}=24.1\pm3.4~{\rm km~s}^{-1}$.

\begin{figure}
 {\includegraphics[angle=0,width=85mm]{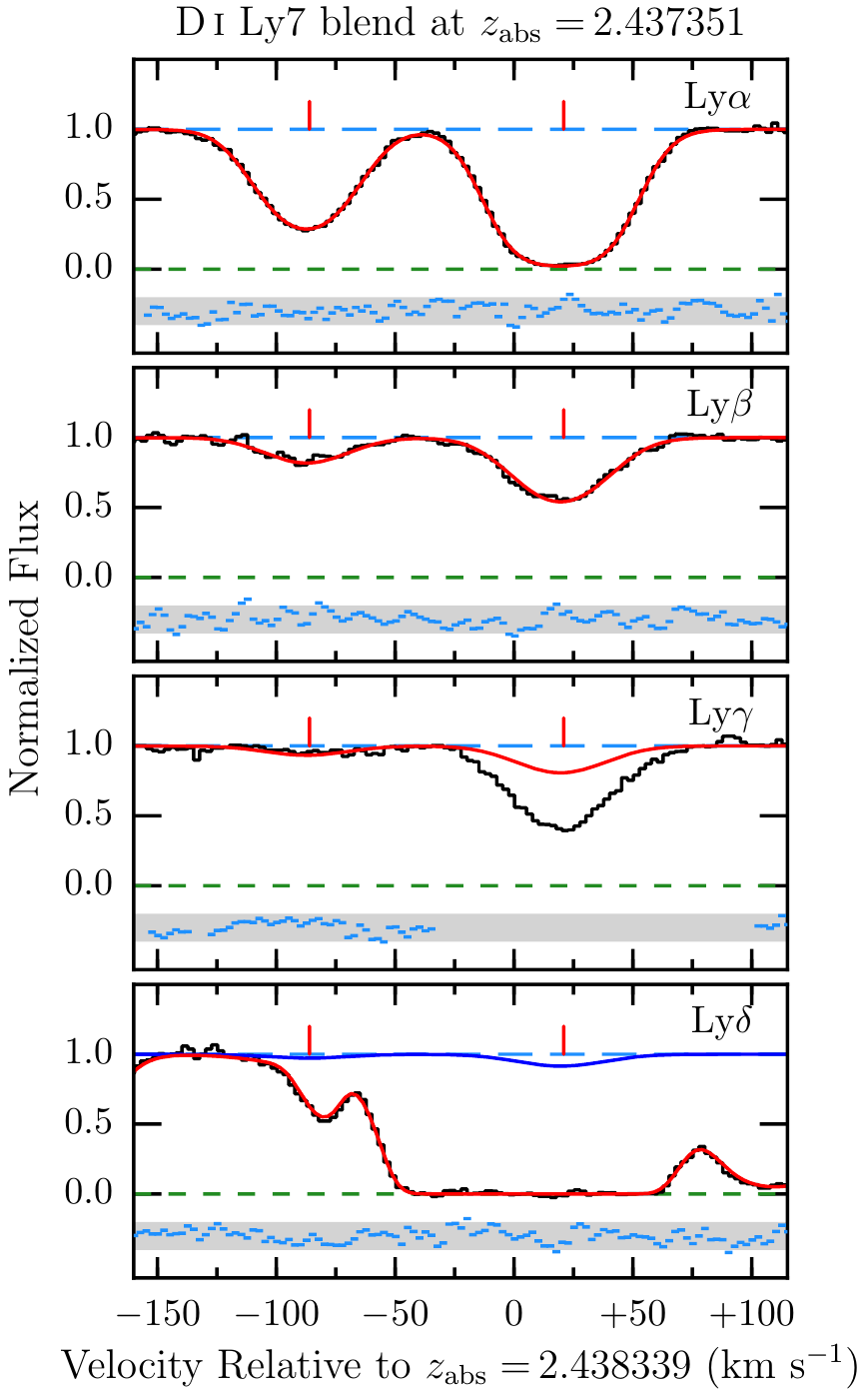}
 \hspace{6mm}\includegraphics[angle=0,width=85mm]{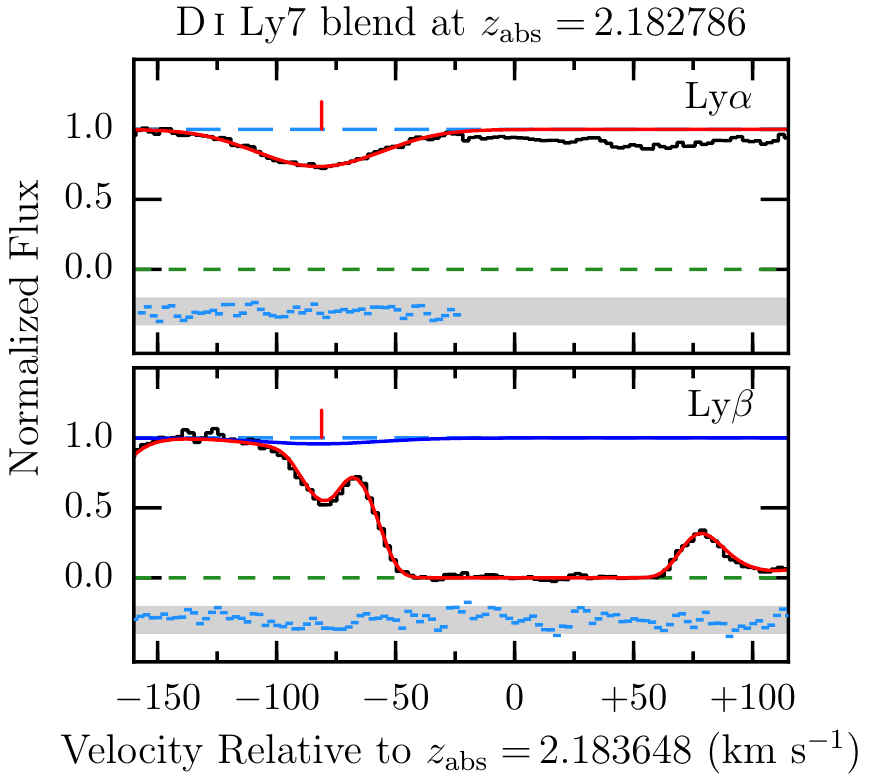}}\\
 \caption{Same as Figures~\ref{fig:blend1} and \ref{fig:blend2}, but illustrating the two blends associated with the \DI\ Ly7 transition of the sub-DLA.}
 \label{fig:blend3}
\end{figure}

Finally, we find that the \DI\ Ly7 line is blended with two separate (weak) absorption systems, which we present in Figure~\ref{fig:blend3}. The first blend is from the Ly$\delta$ transition of a system at $z_{\rm abs}=2.437351\pm0.000002$, which has an \HI\ column density of log$_{10}$~\NHI/cm$^{-2} = 13.594\pm0.005$ and a Doppler parameter $b_{\rm tot}=23.5\pm0.3~{\rm km~s}^{-1}$. The second blend is due to the Ly$\beta$ absorption from a system at $z_{\rm abs}=2.182786\pm0.000006$. The \HI\ column density and Doppler parameter of this system are
log$_{10}$~\NHI/cm$^{-2} = 13.15\pm0.02$ and
$b_{\rm tot}=34.5\pm1.1~{\rm km~s}^{-1}$, respectively.

\end{document}